\documentclass[11pt]{article}
\usepackage[margin=1in]{geometry}
\usepackage{amsmath,amssymb,amsfonts,bm}
\usepackage{booktabs}
\usepackage[hidelinks]{hyperref}
\usepackage{enumitem}
\usepackage{setspace}
\usepackage{mathtools}
\usepackage{titlesec}
\usepackage{microtype}
\usepackage{xcolor}

\titleformat{\section}{\normalfont\large\bfseries}{\thesection}{1em}{}
\titleformat{\subsection}{\normalfont\normalsize\bfseries}{\thesubsection}{1em}{}

\newcommand{\Nstar}{N_{\star}}
\newcommand{\dd}{\mathrm{d}}

\title{\bfseries Biological Time, Evolutionary Optimization, and Gauge
Coherence:\\[2pt]
\large A Thermodynamic Synthesis of the Principle of Biological Time Equivalence}
\author{Mesfin Asfaw Taye}
\date{}

\begin{document}
\maketitle

\begin{abstract}
\noindent Biological theory usually treats time as an external chronological
variable against which growth, aging, and ecological change are parametrized. Yet
living systems also generate an internal measure of duration through physiological
cycling and irreversible entropy production, and the regularities of allometric
lifespan scaling, biological clocks, life-history evolution, ecological
synchronization, and disease are ordinarily studied in isolation rather than within
a single thermodynamic internal-time framework. The Principle of Biological Time
Equivalence (PBTE) proposes such a framework. We define biological proper time by
$\theta_i(t)=\int_0^t f_i(s)\,\dd s$, where $f_i$ is an intrinsic physiological
frequency, so that the lifetime accumulated internal time satisfies
$\theta_i(L_i)\approx N_{\star,i}$, or in the stationary limit
$f_i L_i \approx N_{\star}$. Assigning an entropy cost per biological tick,
$\sigma_{0,i}=\dot\Sigma_i/f_i$, recasts this as an entropy-normalized internal-time
budget, so that aging is the expenditure of finite thermodynamic duration rather
than the passage of calendar time. On this foundation we formulate life-history
evolution as constrained optimization on the PBTE manifold and derive the
scale-free elasticity-balance condition $E_f=E_L$ together with a shadow price of
biological time that rises under extrinsic mortality; we represent ecosystems as
spectra of interacting biological clocks with Kuramoto and Adler synchronization
thresholds; we treat the organism as a coupled-clock ensemble whose temporal
precision carries an entropy cost bounded by the thermodynamic uncertainty
relation; and we express phase freedom as a gauge symmetry with covariant rate
$D_0\theta=\dot\theta-A_0$, deriving the lifespan--rate relation from a conserved
temporal Noether charge. The framework yields a clinical reading of aging, disease,
chronotherapy, cancer, and viral latency as changes in internal-time rate, entropy
cost, or synchronization. The comparative cycle-count regularities of mammals and
birds are the strongest empirical anchors; the ecological, clinical, and
gauge-theoretic developments are testable theoretical predictions that require
independent validation. PBTE thereby reframes lifespan and healthspan as the
expenditure, deformation, and synchronization of a finite biological duration.
\end{abstract}


\section{Introduction}
\label{sec:introduction}

Time is ordinarily introduced into biological theory as an external independent
variable. Populations grow in time, organisms age in time, circadian systems
oscillate in time, cells divide in time, and ecosystems recover from perturbations
in time. In this conventional formulation, time functions as a neutral coordinate:
it measures the interval over which biological change occurs, but it is not itself
generated by the living system. The Principle of Biological Time Equivalence
(PBTE) begins from a different premise. Living systems do not merely evolve within
an externally imposed temporal background; they also generate an intrinsic measure
of duration through their own physiological activity, metabolic turnover,
regulatory cycling, repair, and irreversible entropy production. The biologically
relevant temporal coordinate is therefore not chronological time alone, but an
internally accumulated time constructed from the processes by which life maintains
itself away from equilibrium.

For organism \(i\), we define biological proper time as
\begin{equation}
\theta_i(t)=\int_0^t f_i(s)\,\mathrm{d}s,
\label{eq:theta_intro}
\end{equation}
where \(f_i(t)\) is an intrinsic biological rate. Depending on the scale of
description, \(f_i(t)\) may denote cardiac frequency, respiratory frequency,
cellular turnover, circadian-metabolic cycling, neural firing statistics,
mitochondrial activity, or another coarse-grained physiological clock. Equation
\eqref{eq:theta_intro} is the conceptual foundation of PBTE. It replaces the
external question ``how much clock time has passed?'' with the internal question
``how much biological activity has been accumulated?'' A small mammal with rapid
cardiac and metabolic rhythms advances rapidly through this internal coordinate,
whereas a large mammal with slower rhythms advances more slowly. Thus, biological
duration is not exhausted by calendar time; it is the accumulated path length of
physiological activity.

The terminal biological proper time accumulated over the lifespan \(L_i\) is
\begin{equation}
\theta_i(L_i)=\int_0^{L_i} f_i(t)\,\mathrm{d}t \equiv N_{\star,i}.
\label{eq:Nstar_intro}
\end{equation}
The empirical motivation for PBTE comes from the long-standing observation that,
within broad physiological classes, the product of characteristic physiological
frequency and lifespan is often more narrowly distributed than either variable
separately. In the canonical mammalian case, this appears as the approximate
lifetime cardiac-cycle count, whose order of magnitude is close to \(10^9\)
heartbeats for many reference mammals
\cite{Lindstedt1981,SchmidtNielsen1984,West1997,Brown2004,Hulbert2007}.
In the steady-rate approximation, Eq.~\eqref{eq:Nstar_intro} reduces to
\begin{equation}
f_i L_i = N_\star .
\label{eq:fL_intro}
\end{equation}
Equation \eqref{eq:fL_intro} should not be interpreted as a rigid universal law of
heartbeats. Rather, it is the leading-order projection of a deeper internal-time
constraint: organisms with faster physiological rhythms consume a finite biological
cycle budget over shorter chronological intervals, whereas organisms with slower
rhythms distribute a comparable internal trajectory over longer chronological
intervals.

PBTE becomes a thermodynamic principle when each biological tick is assigned an
irreversible entropy cost. Let \(\dot{\Sigma}_i(t)\) denote the entropy-production
rate of organism \(i\). The entropy cost per physiological cycle is then
\begin{equation}
\sigma_{0,i}(t)
=
\frac{\dot{\Sigma}_i(t)}{f_i(t)}
=
\frac{\mathrm{d}\Sigma_i}{\mathrm{d}\theta_i}.
\label{eq:sigma0_intro}
\end{equation}
This quantity distinguishes two biological trajectories that may possess the same
cycle count but different dissipative histories. A heartbeat, breath, cell cycle,
or metabolic oscillation executed under low physiological stress is not
thermodynamically equivalent to the same nominal cycle executed under inflammation,
thermal strain, oxidative stress, immune activation, or regulatory failure. The
entropy-normalized internal time is therefore defined as
\begin{equation}
\Theta_{\sigma,i}(t)
=
\int_0^t
\frac{\sigma_{0,i}(s)}{\sigma_{0,\mathrm{ref}}}
f_i(s)\,\mathrm{d}s,
\label{eq:Theta_sigma_intro}
\end{equation}
where \(\sigma_{0,\mathrm{ref}}\) is a reference entropy cost per biological cycle.
The corresponding normalized PBTE age is
\begin{equation}
A_{\mathrm{PBTE},i}(t)
=
\frac{\Theta_{\sigma,i}(t)}{N_{\star,\mathrm{ref}}}.
\label{eq:APBTE_intro}
\end{equation}
Equations \eqref{eq:sigma0_intro}--\eqref{eq:APBTE_intro} clarify the physical
meaning of biological aging in the PBTE framework. Aging is not identified with
the passive passage of chronological time. It is the progressive expenditure of an
entropy-weighted internal-time budget. Chronological time, physiological cycle
count, and entropy cost per cycle are therefore distinct quantities. Their
confusion is one reason why organisms of equal chronological age may possess very
different biological ages, and why equal numbers of physiological cycles may have
different consequences depending on the thermodynamic conditions under which those
cycles occur.

This formulation is closely related to, but conceptually distinct from,
classical allometric theory. Metabolic scaling theory explains how physiological
rates, metabolic power, and lifespan vary with body mass and temperature through
regular scaling relations \cite{West1997,Gillooly2001,Brown2004,SchmidtNielsen1984}.
PBTE does not replace this framework. Instead, it interprets the resulting
frequency--lifespan product as an internal-time budget and asks how that budget is
generated, conserved, deformed, or dissipatively modified. In this sense, PBTE
reverses the usual explanatory direction. The approximate invariance of \(f_iL_i\)
is not treated merely as a numerical consequence of fitted allometric exponents;
rather, it is interpreted as the observable projection of a thermodynamic
accounting relation between physiological cycling, entropy production, and
lifespan.

The present paper develops this interpretation into a unified mathematical
framework for biological duration. The framework draws from several established
but largely disconnected literatures. Comparative physiology and allometric
biology provide the empirical basis for scaling relations among body mass,
metabolic rate, heart rate, respiration rate, and lifespan
\cite{SchmidtNielsen1984,West1997,Brown2004,Speakman2005,Hulbert2007}.
The classical rate-of-living tradition, together with modern membrane-pacemaker
and oxidative-damage theories, connects metabolic intensity to longevity while
also revealing important exceptions and clade-specific deviations
\cite{Pearl1928,Harman1956,Barja1998,Hulbert2007,Speakman2005}. Life-history
theory describes how organisms allocate finite energetic and physiological
resources among growth, reproduction, maintenance, and survival
\cite{Stearns1992,Charnov2002,Caswell2001}. Comparative senescence biology
demonstrates that aging is not a single universal trajectory, but a family of
dynamic patterns shaped by ecology, physiology, and evolutionary history
\cite{Finch1990,Jones2014}. Modern biological-age biomarkers, including
DNA-methylation clocks and multi-omic aging measures, provide empirical evidence
that biological age can diverge substantially from chronological age
\cite{Horvath2013,Levine2018,Lu2019,Lu2023}.

In parallel, circadian biology, chronobiology, and the theory of coupled
oscillators show that living systems contain multiple interacting clocks rather
than a single biological timer \cite{Winfree1980,Kuramoto1984,Pikovsky2001,
Panda2021,Asher2015}. Nonequilibrium thermodynamics and stochastic thermodynamics
provide the language required to assign an entropy cost to irreversible biological
processes \cite{Prigogine1967,Seifert2012}. Thermodynamic uncertainty relations
further suggest that the precision of biological timekeeping is constrained by
dissipation \cite{Barato2015,Gingrich2016,Hasegawa2021}. Clinical and ecological
applications also motivate an internal-time formulation: hibernation and torpor
suppress physiological time by reducing metabolic throughput
\cite{Carey2003,Geiser2004}; chronotherapy depends on the phase relation between
drug action and biological clocks \cite{Levi2010,Asher2015}; cancer metabolism
reorganizes cellular energetic and temporal regulation
\cite{Hanahan2011,VanderHeiden2009}; and viral latency may be interpreted as a
partial suspension of replicative internal time \cite{Finzi1997,Chun1997,
Siliciano2003,Deeks2016}. What is missing across these literatures is a single
internal-time variable capable of connecting lifespan scaling, entropy production,
aging, physiological clock coupling, disease progression, ecological
synchronization, and therapeutic timing within one mathematical accounting
principle. Providing that variable is the central objective of this work.

PBTE is therefore not merely the empirical statement that many mammals complete
approximately comparable numbers of heartbeats over their lives. It is a proposed
nonequilibrium principle for biological duration. Its central claim is that living
systems are organized by an entropy-normalized internal-time budget, of which
lifetime cycle count is the simplest observable manifestation. This distinction is
essential. Established comparative biology supports approximate physiological
cycle-count regularities and allometric scaling laws within restricted groups
\cite{Lindstedt1981,SchmidtNielsen1984,West1997,Brown2004,Hulbert2007}. The new
contribution of the present paper is to reinterpret these regularities as
projections of a broader thermodynamic structure and to extend that structure into
aging, evolutionary optimization, ecological synchronization, multi-clock
physiology, and disease.

The present paper has several specific aims. First, it formulates biological
proper time as an internally generated coordinate rather than an externally
imposed parameter. Second, it introduces an entropy-normalized biological age that
distinguishes chronological duration, physiological cycle number, and dissipative
cost per cycle. Third, it embeds PBTE within allometric and metabolic scaling
theory while clarifying why allometric mass cancellation alone is insufficient as
a complete theory of biological duration. Fourth, it develops a variational
life-history formulation in which biological time acts as a constrained resource
allocated among growth, reproduction, repair, maintenance, and survival. Fifth, it
extends PBTE to ecological systems by treating communities as interacting spectra
of biological clocks. Sixth, it develops a multi-clock formulation for coupled
physiological rhythms, including cardiac, respiratory, circadian, metabolic,
immune, and cellular clocks. Seventh, it introduces a geometric and
gauge-invariant language for biological phase, entrainment, temporal mismatch,
and phase resetting. Eighth, it translates the theory into applications in aging,
disease, chronotherapy, healthspan, ecological resilience, and thermodynamic
habits. Ninth, it identifies falsifiable predictions and distinguishes established
comparative physiology from theoretical extrapolation.

The central thesis is that biological time is not merely an index of age. It is a
dynamical coordinate whose conservation, deformation, synchronization, and
misalignment organize living systems across scales. At the organismal scale, PBTE
relates physiological pace to lifespan. At the cellular scale, it links turnover,
repair, damage accumulation, and entropy production. At the ecological scale, it
suggests that communities are structured not only by biomass, abundance, and
trophic interactions, but also by interacting frequency spectra. At the clinical
scale, it distinguishes chronological age from the rate at which entropy-normalized
biological time is consumed. At the therapeutic scale, it suggests that the timing
of intervention should be referenced not only to clock time, but also to internal
phase and dissipative state. At the geometric scale, it motivates a covariant
description of biological phase in which entrainment, desynchronization,
resynchronization, and chronotherapeutic timing can be treated within a common
mathematical language.

The remainder of the paper is organized as follows. Section~\ref{sec:foundation}
establishes the definitions of biological proper time, entropy cost per cycle, and
entropy-normalized PBTE age, and connects them to metabolic scaling and the
conditions under which lifetime cycle-count regularities emerge.
Section~\ref{sec:evolution} formulates life-history evolution as constrained
optimization on the PBTE manifold and derives the corresponding elasticity-balance
and shadow-price conditions. Section~\ref{sec:fitness} anchors the performance
functional in demographic and thermodynamic fitness surfaces. Section~\ref{sec:eco}
extends the framework to ecological synchronization, and Section~\ref{sec:ecosystem}
develops the associated ecosystem entropy throughput and temporal resilience.
Section~\ref{sec:control} treats evolution as control dynamics on the manifold.
Section~\ref{sec:clocks} develops the coupled-clock description of physiological
rhythms and the entropy cost of their coordination. Section~\ref{sec:gauge}
introduces the gauge-invariant formulation of biological phase.
Section~\ref{sec:medicine} applies the theory to aging, disease, chronotherapy, and
healthspan. Section~\ref{sec:implications} states the implications and limitations,
and Section~\ref{sec:open_problems} sets out falsifiable predictions and empirical
tests. The paper closes by clarifying the boundary between established empirical
regularities and the broader theoretical programme proposed here.

The epistemic status of the framework must be stated explicitly. The strongest
empirical support for PBTE lies in comparative physiological regularities among
endothermic vertebrates and in the structured deviations produced by body size,
temperature, torpor, flight, diving, ecological hazard, and clade-specific
metabolic organization. The entropy-normalized aging formulation, the ecological
synchronization theory, the multi-clock oscillator model, and the gauge-theoretic
description of biological phase are theoretical extensions. They are
mathematically motivated and biologically interpretable, but they require
independent empirical validation. Accordingly, the purpose of this paper is not to
declare a completed theory of biological time. Its purpose is to formulate a
falsifiable mathematical programme: to define the internal-time variable, derive
its consequences, state its measurable predictions, and identify the empirical
conditions under which PBTE succeeds or fails.

\section{Foundational Definitions: Biological Proper Time and Entropy-Normalized PBTE}
\label{sec:foundation}

The central object of the PBTE framework is not chronological time itself, but an
internal coordinate generated by the irreversible activity of the organism. For an
organism \(i\), let \(x_i(t)\) denote its physiological state, \(E_i(t)\) its ecological
or environmental context, and \(T_i(t)\) its thermal regime. The effective intrinsic
rate of biological time accumulation is written as
\[
\omega_i(t)\equiv \omega_i\!\left(x_i(t),E_i(t),T_i(t)\right),
\]
and biological proper time is defined by the path integral
\begin{equation}
\theta_i(t)=\int_0^t
\omega_i\!\left(x_i(s),E_i(s),T_i(s)\right)\,\dd s .
\label{eq:proper_time}
\end{equation}
This definition is deliberately state-dependent. A living organism is not a passive
clock with a fixed tick rate. Fever, inflammation, torpor, caloric restriction,
hibernation, diving bradycardia, circadian disruption, and metabolic disease all
alter the physiological state and therefore alter \(\omega_i\). Equation
\eqref{eq:proper_time} consequently defines a biological worldline: two organisms
may experience the same chronological interval while traversing different
distances in internal time.

In the simplest stationary limit, the organism occupies a near-homeostatic state
over the observation interval, so that
\[
\omega_i\!\left(x_i(t),E_i(t),T_i(t)\right)\simeq f_i ,
\]
where \(f_i\) is a characteristic physiological frequency. Then
\begin{equation}
\theta_i(t)\simeq f_i t .
\label{eq:stationary_theta}
\end{equation}
The terminal value of biological proper time over a lifespan \(L_i\) is
\begin{equation}
N_{\star,i}
\equiv
\theta_i(L_i)
=
\int_0^{L_i}
\omega_i\!\left(x_i(t),E_i(t),T_i(t)\right)\,\dd t .
\label{eq:Nstar_general}
\end{equation}
When the organism is approximately stationary, this reduces to
\begin{equation}
N_{\star,i}\simeq f_iL_i .
\label{eq:Nstar_stationary}
\end{equation}
The dimensionless internal-age coordinate is therefore
\begin{equation}
a_i(t)=\frac{\theta_i(t)}{N_{\star,i}} .
\label{eq:bioage}
\end{equation}
In the idealized PBTE limit, \(a_i(0)=0\) and \(a_i(L_i)=1\). The value of
\eqref{eq:bioage} is that it separates calendar age from intrinsic biological
progress. A fast physiological state moves rapidly along the internal coordinate;
a slow physiological state moves slowly. Thus, the same chronological age can
correspond to different fractions of the organism's internal budget.

The preceding definition is kinematic. To give it thermodynamic content, let
\(\dot{\Sigma}_i(t)\ge 0\) denote the irreversible entropy-production rate of the
organism. In an open nonequilibrium system, this entropy production arises from
biochemical reaction networks, ion pumping, molecular repair, biosynthesis,
mechanical work, transport, neural signaling, and heat dissipation
\cite{Prigogine1967,Seifert2012}. During an infinitesimal chronological interval
\(\dd t\), the organism produces
\begin{equation}
\dd\Sigma_i=\dot{\Sigma}_i(t)\,\dd t .
\label{eq:dSigma_dt}
\end{equation}
At the same time, the internal clock advances by
\begin{equation}
\dd\theta_i=\omega_i(t)\,\dd t .
\label{eq:dtheta_dt}
\end{equation}
Eliminating \(\dd t\) between \eqref{eq:dSigma_dt} and \eqref{eq:dtheta_dt} gives
\begin{equation}
\dd\Sigma_i
=
\frac{\dot{\Sigma}_i(t)}{\omega_i(t)}
\,\dd\theta_i .
\label{eq:dSigma_dtheta}
\end{equation}
This identifies the instantaneous entropy cost per biological tick as
\begin{equation}
\sigma_{0,i}(t)
\equiv
\frac{\dot{\Sigma}_i(t)}{\omega_i(t)}
=
\frac{\dd\Sigma_i}{\dd\theta_i}.
\label{eq:sigma0_def}
\end{equation}
Equation~\eqref{eq:sigma0_def} is the thermodynamic bridge between biological time
and irreversible dissipation. A physiological cycle executed under low dissipation
has a small entropy cost; the same nominal cycle executed under inflammation,
oxidative stress, fever, mitochondrial inefficiency, or regulatory failure has a
larger entropy cost. Thus biological time can be counted either as raw cycles or
as entropy-weighted cycles.

Integrating \eqref{eq:dSigma_dtheta} over the lifespan gives
\begin{equation}
\Sigma_i^{\mathrm{life}}
=
\int_0^{L_i}\dot{\Sigma}_i(t)\,\dd t
=
\int_0^{N_{\star,i}}
\sigma_{0,i}(\theta)\,\dd\theta .
\label{eq:Sigma_life_theta}
\end{equation}
The lifetime-average entropy cost per biological tick is
\begin{equation}
\left\langle \sigma_{0,i}\right\rangle
=
\frac{1}{N_{\star,i}}
\int_0^{N_{\star,i}}
\sigma_{0,i}(\theta)\,\dd\theta .
\label{eq:sigma0_average}
\end{equation}
Combining \eqref{eq:Sigma_life_theta} and \eqref{eq:sigma0_average} yields the
PBTE accounting relation
\begin{equation}
N_{\star,i}
=
\frac{\Sigma_i^{\mathrm{life}}}
{\left\langle \sigma_{0,i}\right\rangle}.
\label{eq:Nstar_entropy}
\end{equation}
This equation is exact once \(\theta_i\), \(\Sigma_i\), and \(\sigma_{0,i}\) are
defined. The empirical and physical hypothesis is not the identity itself, but the
claim that \(\left\langle \sigma_{0,i}\right\rangle\), after suitable normalization
for body mass, temperature, and physiological class, is sufficiently constrained
within a clade to make \(N_{\star,i}\) narrowly distributed. In this sense PBTE is
not a statement that organisms die at an exact universal number of heartbeats,
breaths, or metabolic events. It is a statement that biological trajectories are
organized by an entropy-normalized internal-time budget.

A useful operational approximation follows from the open-system entropy balance.
Let \(S_i(t)\) denote the coarse-grained internal entropy of the organism,
\(\dot e_{p,i}(t)\ge0\) the irreversible entropy-production rate, and
\(\dot h_{d,i}(t)\ge0\) the entropy export rate. The open-system second law then
takes the balance form
\begin{equation}
\frac{\dd S_i}{\dd t}
=
\dot e_{p,i}(t)-\dot h_{d,i}(t),
\label{eq:entropy_balance}
\end{equation}
so that the internal entropy changes at the rate of irreversible production minus
the rate of entropy export to the environment. The irreversible production is the
quantity already denoted \(\dot{\Sigma}_i\equiv\dot e_{p,i}\). In adult homeostasis
the organism is maintained near a nonequilibrium steady state, so that
\(\dd S_i/\dd t\simeq0\) over intermediate physiological timescales and production is
balanced by export,
\[
\dot e_{p,i}(t)\simeq \dot h_{d,i}(t).
\]
To leading order, the entropy exported as heat is the dissipated metabolic power
divided by absolute temperature,
\begin{equation}
\dot h_{d,i}(t)\simeq \frac{P_i(t)}{T_i(t)} .
\label{eq:heat_entropy_export}
\end{equation}
Combining the steady-state balance with Eq.~\eqref{eq:heat_entropy_export} gives the
metabolic closure for the entropy-production rate,
\begin{equation}
\dot{\Sigma}_i(t)\equiv\dot e_{p,i}(t)
\simeq \frac{P_i(t)}{T_i(t)} ,
\label{eq:metabolic_closure}
\end{equation}
and the entropy cost per biological tick becomes
\begin{equation}
\sigma_{0,i}(t)
\simeq
\frac{P_i(t)}
{T_i(t)\,\omega_i(t)} .
\label{eq:sigma0_metabolic}
\end{equation}
In mass-normalized form,
\begin{equation}
\sigma_{0,i}^{\ast}(t)
\equiv
\frac{\sigma_{0,i}(t)}{M_i}
\simeq
\frac{P_i(t)}
{M_iT_i(t)\,\omega_i(t)} .
\label{eq:sigma0_mass}
\end{equation}
Equation~\eqref{eq:sigma0_mass} is the measurable closure: it relates the
thermodynamic price of one biological tick to metabolic power, body temperature,
body mass, and physiological frequency.

The entropy-weighted version of biological proper time is then
\begin{equation}
\Theta_{\sigma,i}(t)
=
\int_0^t
\frac{\sigma_{0,i}(s)}{\sigma_{0,\mathrm{ref}}}
\omega_i(s)\,\dd s ,
\label{eq:Theta_sigma}
\end{equation}
where \(\sigma_{0,\mathrm{ref}}\) is a fixed reference entropy cost per cycle. Using
\(\sigma_{0,i}(t)\omega_i(t)=\dot{\Sigma}_i(t)\), Eq.~\eqref{eq:Theta_sigma}
becomes
\begin{equation}
\Theta_{\sigma,i}(t)
=
\frac{1}{\sigma_{0,\mathrm{ref}}}
\int_0^t
\dot{\Sigma}_i(s)\,\dd s
=
\frac{\Sigma_i(t)}{\sigma_{0,\mathrm{ref}}}.
\label{eq:Theta_sigma_entropy}
\end{equation}
Thus entropy-normalized biological time has two equivalent interpretations: it is
the number of entropy-weighted biological ticks, and it is accumulated entropy
production measured in units of the reference entropy cost per tick. The normalized
PBTE age is
\begin{equation}
A_{\mathrm{PBTE},i}(t)
=
\frac{\Theta_{\sigma,i}(t)}
{N_{\star,\mathrm{ref}}}
=
\frac{1}{N_{\star,\mathrm{ref}}}
\int_0^t
\frac{\sigma_{0,i}(s)}{\sigma_{0,\mathrm{ref}}}
\omega_i(s)\,\dd s .
\label{eq:APBTE}
\end{equation}
Defining the reference entropy-cycle budget
\begin{equation}
\Sigma_{\mathrm{ref}}
=
\sigma_{0,\mathrm{ref}}N_{\star,\mathrm{ref}},
\label{eq:Sigma_ref}
\end{equation}
Eq.~\eqref{eq:APBTE} can be written in the compact thermodynamic form
\begin{equation}
A_{\mathrm{PBTE},i}(t)
=
\frac{\Sigma_i(t)}{\Sigma_{\mathrm{ref}}}.
\label{eq:APBTE_entropy}
\end{equation}
This is the most concise expression of the framework: PBTE age is the fraction of
a reference entropy-cycle budget that has been consumed.

Differentiating \eqref{eq:APBTE} gives the instantaneous velocity of biological
aging,
\begin{equation}
\frac{\dd A_{\mathrm{PBTE},i}}{\dd t}
=
\frac{1}{N_{\star,\mathrm{ref}}}
\frac{\sigma_{0,i}(t)}{\sigma_{0,\mathrm{ref}}}
\omega_i(t)
=
\frac{\dot{\Sigma}_i(t)}{\Sigma_{\mathrm{ref}}}.
\label{eq:aging_velocity}
\end{equation}
With the metabolic closure \eqref{eq:metabolic_closure},
\begin{equation}
\frac{\dd A_{\mathrm{PBTE},i}}{\dd t}
\simeq
\frac{P_i(t)}
{T_i(t)\Sigma_{\mathrm{ref}}}.
\label{eq:aging_velocity_metabolic}
\end{equation}
Equations~\eqref{eq:APBTE}--\eqref{eq:aging_velocity_metabolic} clarify the
conceptual distinction that is essential for the remainder of the paper. An
organism can age faster because its physiological clock runs faster, because each
tick is more dissipative, or because both occur simultaneously. Conversely, aging
can slow when physiological rate is reduced without functional collapse, when
maintenance becomes thermodynamically cheaper per tick, or when the accessible
entropy-cycle budget is enlarged. Chronological time is therefore only the
external parameter; biological age is the position of the organism along an
internal thermodynamic trajectory.

The simplest geometric representation of this trajectory is the PBTE manifold.
Let \(f>0\) denote a stationary intrinsic pace and \(L>0\) a lifespan. In the
constant-rate limit, the admissible life histories satisfying the PBTE constraint
form the level set
\begin{equation}
\mathcal{M}_{\mathrm{PBTE}}
=
\bigl\{(f,L)\in\mathbb{R}_{+}^{2}: fL=\Nstar\bigr\}.
\label{eq:manifold}
\end{equation}
In logarithmic coordinates \(u=\log f\) and \(v=\log L\), this becomes
\begin{equation}
u+v=\log \Nstar .
\label{eq:logmanifold}
\end{equation}
The tangent condition is
\begin{equation}
\dd v=-\dd u,
\qquad
\text{or equivalently}
\qquad
\dd\log L=-\dd\log f .
\label{eq:tangent}
\end{equation}
Thus an admissible infinitesimal displacement along the PBTE manifold must satisfy
\begin{equation}
\frac{\dd L}{L}
=
-\frac{\dd f}{f}.
\label{eq:relative_tangent}
\end{equation}
A proportional increase in intrinsic pace is compensated by an equal proportional
decrease in lifespan. The normal direction to the manifold is
\begin{equation}
\dd\log f+\dd\log L\ne0,
\label{eq:normal_deviation}
\end{equation}
and deviations in this direction correspond to changes in the effective budget
\(N_\star\), changes in entropy cost per cycle, clade-specific physiological
multipliers, or pathological departures from the reference regime.

This geometric formulation is important because it converts PBTE from a numerical
rule into a constraint on admissible life-history variation. Evolutionary
variation can move a lineage along the manifold by trading pace against duration.
However, movement normal to the manifold requires a change in the effective
entropy-cycle budget or in the thermodynamic price of biological ticks. Bats,
diving mammals, birds, primates, ectotherms, and disease states should therefore
not be treated as random scatter around \(fL=\Nstar\). They should be analyzed as
structured deformations of the reference manifold:
\begin{equation}
f_iL_i
=
\Phi_i\,\Nstar ,
\label{eq:multiplier_manifold}
\end{equation}
where \(\Phi_i\) is a dimensionless multiplier encoding physiological mechanisms
such as duty-cycle suppression, thermal normalization, neural investment,
mitochondrial efficiency, altered repair, or chronic hypermetabolic stress.
Taking logarithms gives
\begin{equation}
\log f_i+\log L_i
=
\log \Nstar+\log\Phi_i .
\label{eq:log_multiplier}
\end{equation}
Thus clade-specific mechanisms appear geometrically as parallel shifts of the
reference PBTE line in log-space.

The remainder of the paper builds on these definitions. Evolutionary optimization
is treated as constrained motion on \(\mathcal{M}_{\mathrm{PBTE}}\); ecological
synchronization is treated as coupling among species-specific biological-time
manifolds; multi-clock physiology is treated as synchronization among internal
oscillators with distinct \(\theta_a(t)\); and the gauge formulation arises because
the origin and local parametrization of biological phase are arbitrary, whereas
the covariant internal-time increment is physically meaningful. These later
developments all depend on the same foundational distinction established here:
biological time is not elapsed calendar time, but an entropy-weighted path length
generated by irreversible physiological activity.

\section{Evolution as Optimization on the PBTE Manifold}
\label{sec:evolution}

A central implication of PBTE is that evolution does not optimize life history in an
unconstrained trait space. Classical life-history theory has long recognized that
growth, reproduction, maintenance, and survival compete for finite energetic and
temporal resources \cite{Stearns1992,Charnov2002,Caswell2001}. PBTE sharpens this
principle by identifying the relevant temporal constraint explicitly: a lineage does
not merely allocate energy across calendar time; it allocates a finite budget of
intrinsic biological time. In the stationary approximation this budget is expressed
by
\begin{equation}
fL=\Nstar,
\label{eq:pbte_constraint_evol}
\end{equation}
where \(f\) is an intrinsic physiological pace, \(L\) is lifespan, and \(\Nstar\) is
the effective lifetime cycle budget after the appropriate clade and physiological
corrections have been applied. Thus, an evolutionary increase in pace cannot be
treated as an independent gain. Unless accompanied by an increase in the effective
budget, it must be paid for by a reduction in lifespan. Conversely, an extension of
lifespan requires a compensating reduction in pace or an enlargement of the
entropy-normalized cycle budget.

Let
\begin{equation}
P(f,L;\eta)>0
\end{equation}
denote a smooth evolutionary performance functional. Depending on the context,
\(P\) may represent lifetime reproductive output, intrinsic population growth rate,
survival-weighted fecundity, or a coarse-grained measure of ecological performance.
The parameter vector \(\eta\) collects environmental and physiological conditions:
resource abundance, extrinsic mortality, predation hazard, reproductive timing,
developmental schedule, body temperature, repair efficiency, and maintenance cost.
No separability of \(P\) is assumed. The only mathematical assumptions required here
are that \(P\in C^2\) on the biologically relevant domain, that an interior optimum
exists, and that \(\Nstar\) is approximately fixed on the evolutionary timescale under
consideration. The PBTE-constrained evolutionary problem is therefore
\begin{equation}
\max_{f,L>0} P(f,L;\eta)
\qquad
\text{subject to}
\qquad
C(f,L)=fL-\Nstar=0 .
\label{eq:opt}
\end{equation}
The corresponding Lagrangian is
\begin{equation}
\mathcal{L}(f,L,\lambda)
=
P(f,L;\eta)-\lambda\,(fL-\Nstar),
\label{eq:lagrangian}
\end{equation}
where \(\lambda\) is the multiplier associated with the intrinsic-time constraint.
Taking first variations gives
\begin{align}
\frac{\partial \mathcal{L}}{\partial f}=0
&\quad\Rightarrow\quad
P_f=\lambda L,
\label{eq:stationarity_f}
\\
\frac{\partial \mathcal{L}}{\partial L}=0
&\quad\Rightarrow\quad
P_L=\lambda f,
\label{eq:stationarity_L}
\\
\frac{\partial \mathcal{L}}{\partial \lambda}=0
&\quad\Rightarrow\quad
fL=\Nstar .
\label{eq:stationarity_constraint}
\end{align}
Equations~\eqref{eq:stationarity_f}--\eqref{eq:stationarity_constraint} state that
at the optimum the gradient of performance is parallel to the gradient of the PBTE
constraint. Evolutionary change can no longer follow the steepest ascent of \(P\) in
the full \((f,L)\) plane; it can only follow the component of that ascent tangent to
the curve \(fL=\Nstar\).

Eliminating \(\lambda\) from
\eqref{eq:stationarity_f} and \eqref{eq:stationarity_L} gives the PBTE Euler
condition
\begin{equation}
\frac{P_f}{P_L}=\frac{L}{f},
\qquad
\text{or equivalently}
\qquad
fP_f=LP_L .
\label{eq:euler}
\end{equation}
This equation is the first central result of the evolutionary theory. It does not
say that the unconstrained performance surface is maximized. Rather, it says that
the observed operating point is the point on the PBTE manifold at which the
weighted marginal return of increasing physiological pace equals the weighted
marginal return of increasing lifespan. A lineage therefore sits at an evolutionary
compromise: accelerating life becomes favorable only until the performance gained
from pace equals the performance lost through compressed duration.

The same condition becomes more transparent when written in logarithmic
elasticities. Define
\begin{equation}
E_f
=
\frac{\partial \log P}{\partial \log f}
=
\frac{fP_f}{P},
\qquad
E_L
=
\frac{\partial \log P}{\partial \log L}
=
\frac{LP_L}{P}.
\label{eq:elasticities}
\end{equation}
Dividing \eqref{eq:euler} by \(P\) gives
\begin{equation}
E_f=E_L .
\label{eq:balance}
\end{equation}
Equation~\eqref{eq:balance} is the elasticity-balance law. It is the scale-free
form of the PBTE evolutionary optimum. At the constrained optimum, a one-percent
increase in intrinsic pace and a one-percent increase in lifespan would produce
the same fractional gain in performance. This form is especially useful because it
does not depend on the units used to measure \(f\), \(L\), or \(P\). It can therefore
be compared across species with vastly different body sizes, heart rates, lifespans,
and ecological settings. In this sense, PBTE converts the pace--duration trade-off
into a dimensionless marginal-equality principle.

The same result can be derived by reducing the constrained problem to one variable.
Since \(L=\Nstar/f\), define the reduced performance
\begin{equation}
\widetilde P(f)
=
P\!\left(f,\frac{\Nstar}{f};\eta\right).
\label{eq:reducedP}
\end{equation}
Differentiating along the PBTE manifold,
\begin{align}
\frac{\dd \widetilde P}{\dd f}
&=
P_f
+
P_L
\frac{\dd}{\dd f}\left(\frac{\Nstar}{f}\right)
\nonumber\\
&=
P_f
-
P_L\frac{\Nstar}{f^2}
\nonumber\\
&=
P_f
-
P_L\frac{L}{f}.
\label{eq:reduced_deriv}
\end{align}
The stationary condition \(\dd\widetilde P/\dd f=0\) again gives
\(fP_f=LP_L\). In logarithmic coordinates,
\begin{equation}
u=\log f,
\qquad
v=\log L,
\qquad
u+v=\log\Nstar .
\end{equation}
Substituting \(v=\log\Nstar-u\), the reduced log-performance becomes
\begin{equation}
\Psi(u)
=
\log P\!\left(e^u,e^{\log\Nstar-u};\eta\right).
\label{eq:Psi}
\end{equation}
Its derivative is
\begin{equation}
\frac{\dd\Psi}{\dd u}
=
\frac{\partial\log P}{\partial\log f}
-
\frac{\partial\log P}{\partial\log L}
=
E_f-E_L .
\label{eq:Phiprime}
\end{equation}
Thus the sign of \(E_f-E_L\) tells the direction of evolutionary motion along the
PBTE manifold. If \(E_f>E_L\), the marginal benefit of pace exceeds the marginal
benefit of duration and the optimum shifts toward faster, shorter lives. If
\(E_f<E_L\), the reverse occurs and the optimum shifts toward slower, longer lives.
The equilibrium is precisely the point where these two elasticities coincide.

Local stability requires more than stationarity. The constrained optimum is stable
only if the reduced log-performance is concave at the stationary point:
\begin{equation}
\Psi''(u^\ast)<0,
\qquad
\Psi''(u)=\frac{\dd}{\dd u}(E_f-E_L).
\label{eq:secondorder_log}
\end{equation}
Biologically, this condition means that the advantage of increasing pace must
decline as pace increases. If faster living initially improves performance through
earlier reproduction, faster development, or more rapid exploitation of resources,
that advantage must eventually be opposed by a cost: accelerated damage, reduced
maintenance, higher mortality exposure, shorter reproductive duration, or greater
entropy production per unit function. Without such curvature, no interior
life-history optimum exists; selection would drive the system to a boundary.

The same stability condition can be expressed directly in \((f,L)\). From
\(\widetilde P(f)=P(f,\Nstar/f)\), a second differentiation gives
\begin{equation}
\frac{\dd^2\widetilde P}{\dd f^2}
=
P_{ff}
-
2\frac{L}{f}P_{fL}
+
\frac{L^2}{f^2}P_{LL}
+
\frac{2L}{f^2}P_L .
\label{eq:d2P}
\end{equation}
A locally stable constrained maximum requires
\begin{equation}
\frac{\dd^2\widetilde P}{\dd f^2}<0 .
\label{eq:constrained_stability}
\end{equation}
The terms in \eqref{eq:d2P} have distinct meanings. The first three terms are the
curvature of the biological performance surface itself. The final term arises
because the PBTE constraint is hyperbolic: moving along \(fL=\Nstar\) bends the
trajectory through the \((f,L)\) plane. Thus stability is not determined only by
the intrinsic curvature of \(P\). It is determined jointly by the fitness surface
and by the geometry of the conserved-time constraint. This point is important:
PBTE can generate a stable constrained optimum even when the unconstrained
performance landscape has no biologically meaningful maximum in the full trait
space.

The Lagrange multiplier \(\lambda\) has a direct biological interpretation. From
\eqref{eq:stationarity_f}--\eqref{eq:stationarity_L},
\begin{equation}
\lambda
=
\frac{P_f}{L}
=
\frac{P_L}{f}.
\label{eq:lambda_stationary}
\end{equation}
Standard constrained-optimization theory gives
\begin{equation}
\lambda
=
\frac{\partial P^\ast}{\partial \Nstar},
\label{eq:shadow}
\end{equation}
where \(P^\ast\) is the optimized performance. Thus \(\lambda\) is the shadow price
of biological time: the marginal increase in optimal performance produced by a
marginal relaxation of the intrinsic-time budget. A large \(\lambda\) means that
biological time is scarce; an additional unit of intrinsic time would substantially
increase performance. A small \(\lambda\) means that the organism is not strongly
limited by intrinsic-time availability under the current ecological regime.

This interpretation allows PBTE to make a sharp ecological prediction. In
environments with high extrinsic mortality, late-life reproduction is discounted
because many individuals die before reaching advanced ages. The elasticity \(E_L\)
therefore decreases relative to \(E_f\). To restore the balance
\(E_f=E_L\), the optimum shifts toward larger \(f^\ast\) and smaller \(L^\ast\).
Dangerous environments select fast life histories not because organisms possess
more biological time, but because the marginal value of spending that time early
exceeds the marginal value of preserving it for later. The shadow price of
biological time rises, and the lineage moves along the PBTE manifold toward a
faster, shorter strategy.

This result can be derived explicitly from a simple demographic performance
surface. Let performance be survival-weighted reproduction,
\begin{equation}
P(f,L)
=
\int_0^L b(t;f)S(t;f)\,\dd t,
\label{eq:demographicP}
\end{equation}
where \(b(t;f)\) is age-specific fecundity and \(S(t;f)\) is survival probability.
Let the total hazard be
\begin{equation}
\mu(t;f)=\mu_{\mathrm{ext}}+\mu_{\mathrm{int}}(t;f),
\label{eq:hazard}
\end{equation}
so that
\begin{equation}
S(t;f)
=
\exp\!\left[-\int_0^t\mu(s;f)\,\dd s\right].
\label{eq:survival}
\end{equation}
A minimal intrinsic hazard model is
\begin{equation}
\mu_{\mathrm{int}}(t;f)
=
\mu_0+\alpha f^\eta t^\gamma,
\qquad
\alpha>0,\quad \eta>0,\quad \gamma\ge0,
\label{eq:intrinsic_hazard}
\end{equation}
where \(f^\eta\) represents the damage or maintenance cost of faster physiological
pace. A simple fecundity scaling is
\begin{equation}
b(t;f)=b_0 f^\rho \chi(t),
\qquad
\rho>0,
\label{eq:fecundity}
\end{equation}
where \(\chi(t)\) encodes reproductive scheduling. Substituting
\eqref{eq:intrinsic_hazard} and \eqref{eq:fecundity} into
\eqref{eq:demographicP}, and imposing \(L=\Nstar/f\), gives the PBTE-reduced
performance
\begin{equation}
\widetilde P(f)
=
\int_0^{\Nstar/f}
b_0 f^\rho \chi(t)
\exp\!\left[
-\int_0^t
\left(
\mu_{\mathrm{ext}}+\mu_0+\alpha f^\eta s^\gamma
\right)\dd s
\right]\dd t .
\label{eq:reduced_demographicP}
\end{equation}
Equation~\eqref{eq:reduced_demographicP} displays the trade-off explicitly. Higher
\(f\) can increase early fecundity through the factor \(f^\rho\), but it shortens the
reproductive window through the upper limit \(\Nstar/f\) and increases intrinsic
hazard through \(\alpha f^\eta\). The optimum is not a universal pace; it is the
pace at which the early benefits of physiological acceleration are exactly balanced
by the loss of lifespan and the accumulation of risk.

A more compact thermodynamic survival kernel reaches the same conclusion. Suppose
that the cumulative survival cost over a life history is approximated by
\begin{equation}
\mathcal{D}(f,L)
=
\kappa_1 f^\eta L+\kappa_2\mu_{\mathrm{ext}}^\zeta L,
\qquad
\kappa_1,\kappa_2>0,
\label{eq:Dkernel}
\end{equation}
where the first term is pace-dependent internal wear and the second is
environmental mortality pressure. Let
\begin{equation}
P_{\mathrm{surv}}(f,L)=\exp[-\mathcal{D}(f,L)] .
\label{eq:Psurv}
\end{equation}
Imposing \(L=\Nstar/f\) gives
\begin{equation}
\widetilde{\mathcal{D}}(f)
=
\kappa_1\Nstar f^{\eta-1}
+
\kappa_2\mu_{\mathrm{ext}}^\zeta\Nstar f^{-1}.
\label{eq:Dreduced}
\end{equation}
Maximizing survival is equivalent to minimizing \(\widetilde{\mathcal{D}}\). The
stationary condition is
\begin{equation}
\frac{\dd\widetilde{\mathcal{D}}}{\dd f}
=
\kappa_1\Nstar(\eta-1)f^{\eta-2}
-
\kappa_2\mu_{\mathrm{ext}}^\zeta\Nstar f^{-2}
=0 .
\label{eq:Dderiv}
\end{equation}
For \(\eta>1\), this gives
\begin{equation}
f^\ast
=
\left[
\frac{\kappa_2\mu_{\mathrm{ext}}^\zeta}
{\kappa_1(\eta-1)}
\right]^{1/\eta},
\label{eq:fstar}
\end{equation}
and therefore
\begin{equation}
L^\ast
=
\frac{\Nstar}{f^\ast}
\propto
\mu_{\mathrm{ext}}^{-\zeta/\eta}.
\label{eq:Lstar}
\end{equation}
Equations~\eqref{eq:fstar}--\eqref{eq:Lstar} give the promised result in explicit
form: higher extrinsic mortality selects a faster optimal pace and a shorter
optimal lifespan. This is not imposed verbally; it follows from PBTE combined with
a survival kernel that penalizes both internal wear and external hazard.

The thermodynamic interpretation of the entire construction is obtained by
returning to entropy production. Let the irreversible entropy-production rate be
\begin{equation}
\dot{\Sigma}(t)=\sigma_0(t)f(t).
\label{eq:entropy_rate_evol}
\end{equation}
If the entropy cost per biological tick is approximately constant along a
reference life history, \(\sigma_0(t)\simeq\sigma_0\), then
\begin{equation}
\Sigma^{\mathrm{life}}
=
\int_0^L \dot{\Sigma}(t)\,\dd t
\simeq
\sigma_0\int_0^L f(t)\,\dd t
=
\sigma_0\Nstar .
\label{eq:life_dissipation}
\end{equation}
In the constant-rate case this becomes
\begin{equation}
\Sigma^{\mathrm{life}}
\simeq
\sigma_0 f\frac{\Nstar}{f}
=
\sigma_0\Nstar .
\label{eq:life_dissipation_constant}
\end{equation}
Thus, among strategies sharing the same \(\Nstar\) and \(\sigma_0\), fast-short and
slow-long life histories dissipate approximately the same total entropy over a
complete lifetime. What differs is the schedule of dissipation. Evolution does not
choose whether dissipation occurs; it chooses when dissipation is spent, how it is
coupled to reproduction, and how much functional performance is obtained per unit
irreversible cost.

This observation is the conceptual bridge between PBTE and life-history evolution.
A fast life history front-loads entropy production, reproduction, and mortality
risk. A slow life history spreads the same intrinsic-time budget over a longer
calendar interval, usually requiring improved maintenance, lower extrinsic hazard,
or lower entropy cost per tick. Exceptional longevity therefore does not represent
escape from thermodynamic accounting. It represents deformation of the accounting
parameters: a reduced effective pace, a lower entropy cost per cycle, a larger
accessible budget, or a clade-specific multiplier:
\begin{equation}
f_iL_i=\Phi_i\Nstar .
\label{eq:evol_multiplier}
\end{equation}
Here \(\Phi_i>1\) indicates an expanded effective internal-time budget or a reduced
entropy cost per biological tick; \(\Phi_i<1\) indicates compression of the
effective budget, elevated dissipation, or high unbuffered hazard. In evolutionary
terms, \(\Phi_i\) is not a fitting constant alone. It is a mechanistic summary of
how a lineage modifies the PBTE constraint.

The implications are broad. First, PBTE gives life-history theory a geometric
constraint: viable strategies lie near a manifold rather than filling the entire
pace--lifespan plane. Second, it gives a marginal condition for evolutionary
equilibrium, \(E_f=E_L\), which can be tested if performance surfaces are estimated
from demographic data. Third, it predicts how ecological pressure shifts the
optimum: higher extrinsic mortality moves lineages toward higher pace and shorter
lifespan, while protected niches allow selection to favor slower pace and longer
duration. Fourth, it gives a thermodynamic interpretation of exceptional longevity:
long-lived clades must either slow their internal clock, lower the entropy cost of
each tick, improve repair sufficiently to enlarge the effective budget, or reduce
external hazard enough that delayed reproduction retains high marginal value.

The same framework also has practical applications. In comparative biology, it
suggests that species should be compared not only by body mass and lifespan but by
their position on the PBTE manifold and by the multiplier \(\Phi_i\) that measures
normal displacement from the reference line. In ecology, it predicts that
disturbed environments should favor fast, short-lived taxa because extrinsic risk
raises the value of early performance. In conservation biology, the loss of
slow-lived species is not merely a loss of biomass or diversity; it is a loss of
low-frequency temporal structure from the ecosystem. In medicine and aging
research, the same mathematics implies that interventions can extend healthy
lifespan either by reducing physiological pace, lowering entropy cost per cycle,
or increasing the resilience of the damage--repair system. The evolutionary
optimization problem therefore becomes the macroscopic counterpart of the clinical
PBTE problem: both ask how a finite entropy-normalized biological-time budget is
spent, protected, accelerated, or prolonged.
\section{Operational Fitness Surfaces}
\label{sec:fitness}

The variational formulation of Section~\ref{sec:evolution} is intentionally
abstract: it states that evolution optimizes performance on the PBTE manifold, but
it does not yet specify what performance means. To obtain testable predictions,
the functional \(P(f,L;\bm{\eta})\) must be anchored to measurable life-history
quantities. The purpose of this section is to show how such anchoring can be done.
We develop two complementary representations. The first is demographic and begins
from survival-weighted reproduction. The second is thermodynamic and begins from
the cost of sustaining a physiological pace under internal wear and external
hazard. The two constructions are not competitors; they are different
coarse-grainings of the same biological trade-off. The demographic surface is
closer to field data, whereas the thermodynamic survival kernel exposes scaling
laws that can be compared across taxa and environments.

Let \(b(t;f)\) denote age-specific fecundity for an organism or lineage operating
at intrinsic pace \(f\), and let \(S(t;f)\) denote survival to age \(t\). The most
direct population-level performance functional is the expected lifetime
reproductive output,
\begin{equation}
P(f,L)=R_0(f,L)
=
\int_0^L b(t;f)S(t;f)\,\dd t .
\label{eq:R0}
\end{equation}
This expression is standard in life-history theory: fitness is increased by
earlier reproduction, greater fecundity, and longer survival, but these quantities
are not independent \cite{Stearns1992,Caswell2001,Charnov2002}. PBTE enters
through the fact that \(L\) is not freely adjustable once the intrinsic pace is
chosen. In the reference regime,
\begin{equation}
L=\frac{\Nstar}{f}.
\label{eq:L_constraint_fitness}
\end{equation}
Thus any increase in pace that raises early fecundity simultaneously shortens the
available reproductive interval.

Survival is determined by a total hazard,
\begin{equation}
\mu(t;f)=\mu_{\mathrm{ext}}+\mu_{\mathrm{int}}(t;f),
\label{eq:hazard_decomp}
\end{equation}
where \(\mu_{\mathrm{ext}}\) represents extrinsic mortality imposed by predation,
accident, infection, starvation, environmental instability, or resource failure,
whereas \(\mu_{\mathrm{int}}\) represents intrinsic mortality generated by damage,
maintenance failure, or physiological wear. The corresponding survival function is
\begin{equation}
S(t;f)
=
\exp\!\left[
-\int_0^t \mu(s;f)\,\dd s
\right].
\label{eq:survival_fs}
\end{equation}
A minimal pace-dependent intrinsic hazard is
\begin{equation}
\mu_{\mathrm{int}}(t;f)
=
\mu_0+\alpha f^\nu t^\gamma,
\qquad
\alpha>0,\quad \nu>0,\quad \gamma\ge0.
\label{eq:muint}
\end{equation}
Here \(\mu_0\) is a baseline intrinsic hazard, \(\alpha\) measures the strength of
pace-induced damage, \(\nu\) is the elasticity of damage with respect to
physiological pace, and \(\gamma\) controls the age-dependence of the intrinsic
hazard. The exponent \(\nu\) allows the cost of speed to be nonlinear. When
\(\nu>1\), fast life becomes increasingly expensive; when \(0<\nu<1\), the cost of
speed is sublinear.

Substituting \eqref{eq:muint} into \eqref{eq:survival_fs} gives an explicit survival
function. Since
\begin{equation}
\int_0^t
\left(
\mu_{\mathrm{ext}}+\mu_0+\alpha f^\nu s^\gamma
\right)\dd s
=
(\mu_{\mathrm{ext}}+\mu_0)t
+
\frac{\alpha f^\nu}{\gamma+1}t^{\gamma+1},
\label{eq:hazard_integral}
\end{equation}
one obtains
\begin{equation}
S(t;f)
=
\exp\!\left[
-(\mu_{\mathrm{ext}}+\mu_0)t
-
\frac{\alpha f^\nu}{\gamma+1}t^{\gamma+1}
\right].
\label{eq:survival_explicit}
\end{equation}
The survival cost therefore contains two distinct clocks. The extrinsic term grows
linearly in chronological exposure time, while the intrinsic term grows with both
chronological time and biological pace. A high value of \(f\) can be beneficial if
it accelerates growth and reproduction, but it also increases the intrinsic hazard
through \(f^\nu\).

To represent the reproductive benefit of fast pace, suppose that fecundity can be
written in the form
\begin{equation}
b(t;f)=b_0 f^\rho \chi(t),
\qquad
\rho>0,
\label{eq:fecundity_fs}
\end{equation}
where \(b_0\) is a scale factor and \(\chi(t)\) describes the age schedule of
reproductive competence. The exponent \(\rho\) captures the advantage of rapid
development, early maturation, or high metabolic throughput. Combining
\eqref{eq:R0}, \eqref{eq:survival_explicit}, \eqref{eq:fecundity_fs}, and
\eqref{eq:L_constraint_fitness}, the PBTE-reduced demographic performance becomes
\begin{equation}
\widetilde P(f)
=
\int_0^{\Nstar/f}
b_0 f^\rho \chi(t)
\exp\!\left[
-(\mu_{\mathrm{ext}}+\mu_0)t
-
\frac{\alpha f^\nu}{\gamma+1}t^{\gamma+1}
\right]
\dd t .
\label{eq:reducedR0}
\end{equation}
Equation~\eqref{eq:reducedR0} is the demographic expression of the PBTE
life-history trade-off. The factor \(f^\rho\) rewards speed. The upper limit
\(\Nstar/f\) penalizes speed by shortening the reproductive window. The exponential
penalizes both ecological exposure and intrinsic wear. The optimum of
\eqref{eq:reducedR0} is therefore not imposed by assumption; it is the point at
which the fecundity gain of faster living, the survival loss from accumulated
hazard, and the geometric constraint \(fL=\Nstar\) balance.

The same conclusion can be expressed in a simpler thermodynamic survival kernel.
Instead of modeling fecundity and survival separately, one may write the survival
performance as the exponential of a cumulative cost:
\begin{equation}
P_{\mathrm{surv}}(f,L)
=
\exp[-\mathcal{C}(f,L)] .
\label{eq:Psurv_cost}
\end{equation}
A minimal cost functional that contains the two relevant penalties is
\begin{equation}
\mathcal{C}(f,L)
=
\kappa_1 f^\nu L
+
\kappa_2\mu_{\mathrm{ext}}^\zeta L,
\qquad
\kappa_1,\kappa_2>0,\quad \zeta>0 .
\label{eq:kernel_cost}
\end{equation}
The first term represents internal wear: high physiological pace sustained over a
lifetime increases damage, repair demand, entropy production, or mortality risk.
The second term represents extrinsic exposure: the longer the organism remains in
a dangerous environment, the more it accumulates risk from causes that are not
directly determined by intrinsic physiology. The parameter \(\zeta\) allows
nonlinear dependence on extrinsic hazard.

Imposing the PBTE constraint \(L=\Nstar/f\), the reduced cost is
\begin{equation}
\widetilde{\mathcal{C}}(f)
=
\kappa_1\Nstar f^{\nu-1}
+
\kappa_2\mu_{\mathrm{ext}}^\zeta \Nstar f^{-1}.
\label{eq:reduced_cost}
\end{equation}
The two terms pull in opposite directions. The internal-wear term increases with
\(f\) when \(\nu>1\), because faster physiology becomes disproportionately costly.
The extrinsic-exposure term decreases with \(f\), because faster life compresses
the calendar time during which external mortality can act. The optimum is found by
minimizing \(\widetilde{\mathcal{C}}\):
\begin{equation}
\frac{\dd \widetilde{\mathcal{C}}}{\dd f}
=
\kappa_1\Nstar(\nu-1)f^{\nu-2}
-
\kappa_2\mu_{\mathrm{ext}}^\zeta \Nstar f^{-2}
=0 .
\label{eq:cost_derivative}
\end{equation}
For \(\nu>1\), multiplication by \(f^2/\Nstar\) yields
\begin{equation}
\kappa_1(\nu-1)f^\nu
=
\kappa_2\mu_{\mathrm{ext}}^\zeta .
\label{eq:fstar_balance}
\end{equation}
Therefore the optimal physiological pace is
\begin{equation}
f^\ast
=
\left[
\frac{\kappa_2\mu_{\mathrm{ext}}^\zeta}
{\kappa_1(\nu-1)}
\right]^{1/\nu}.
\label{eq:fstar_fs}
\end{equation}
The corresponding lifespan is
\begin{equation}
L^\ast
=
\frac{\Nstar}{f^\ast}
=
\Nstar
\left[
\frac{\kappa_1(\nu-1)}
{\kappa_2\mu_{\mathrm{ext}}^\zeta}
\right]^{1/\nu}.
\label{eq:Lstar_fs}
\end{equation}
Thus
\begin{equation}
f^\ast\propto \mu_{\mathrm{ext}}^{\zeta/\nu},
\qquad
L^\ast\propto \mu_{\mathrm{ext}}^{-\zeta/\nu}.
\label{eq:mortality_scaling}
\end{equation}
This scaling is the quantitative heart of the ecological prediction. Hazardous
environments select a faster biological pace and a shorter lifespan because the
cost of remaining exposed becomes large. Protected environments permit slower
pace and longer duration because delayed reproduction and prolonged maintenance
retain value. PBTE therefore recovers a classical life-history result from a
specific thermodynamic-geometric constraint: fast, short lives are favored when
external mortality is high; slow, long lives are favored when external mortality is
low \cite{Stearns1992,Charnov2002}.

A second scaling follows when one asks how the optimal lifespan depends on the
effective intrinsic-time budget itself. Consider a cost functional with an internal
pace penalty and an additional longevity penalty,
\begin{equation}
\mathcal{C}(f,L)
=
\kappa_1 f^\nu L+\kappa_3 L^\xi,
\qquad
\kappa_1,\kappa_3>0,\quad \nu>1,\quad \xi>0.
\label{eq:longevity_cost}
\end{equation}
The term \(\kappa_3L^\xi\) represents the cumulative cost of long duration: repair
burden, exposure to rare events, late-life physiological fragility, or the
increasing difficulty of maintaining regulatory integrity over long times. Using
\(f=\Nstar/L\), the reduced cost as a function of lifespan is
\begin{equation}
\widehat{\mathcal{C}}(L)
=
\kappa_1 \left(\frac{\Nstar}{L}\right)^\nu L
+
\kappa_3 L^\xi
=
\kappa_1\Nstar^\nu L^{1-\nu}
+
\kappa_3 L^\xi .
\label{eq:L_reduced_cost}
\end{equation}
Differentiating,
\begin{equation}
\frac{\dd \widehat{\mathcal{C}}}{\dd L}
=
\kappa_1\Nstar^\nu(1-\nu)L^{-\nu}
+
\kappa_3\xi L^{\xi-1}.
\label{eq:L_derivative}
\end{equation}
Setting this derivative to zero gives
\begin{equation}
\kappa_3\xi L^{\xi-1}
=
\kappa_1(\nu-1)\Nstar^\nu L^{-\nu}.
\label{eq:L_balance}
\end{equation}
Therefore
\begin{equation}
L^{\nu+\xi-1}
=
\frac{\kappa_1(\nu-1)}{\kappa_3\xi}\Nstar^\nu,
\label{eq:L_power}
\end{equation}
and the optimal lifespan scales as
\begin{equation}
L^\ast
=
\left[
\frac{\kappa_1(\nu-1)}{\kappa_3\xi}
\right]^{1/(\nu+\xi-1)}
\Nstar^{\nu/(\nu+\xi-1)} .
\label{eq:L_Nstar_scaling}
\end{equation}
Equation~\eqref{eq:L_Nstar_scaling} supplies an explicit allometric bridge between
lifespan and the effective intrinsic-time budget. The exponent
\[
\frac{\nu}{\nu+\xi-1}
\]
lies below, at, or near unity depending on the strength of the longevity penalty.
When \(\xi\simeq1\), lifespan is approximately proportional to the accessible
cycle budget. When \(\xi>1\), maintaining a long life becomes increasingly costly
and the scaling becomes sublinear. Thus, PBTE predicts not merely that larger
cycle budgets extend lifespan, but that the magnitude of the extension depends on
the curvature of the maintenance-cost surface.

The demographic and thermodynamic surfaces can be unified conceptually. In the
demographic model, the trade-off appears through fecundity, survival, and hazard.
In the thermodynamic model, the same trade-off appears through internal wear and
external exposure. The bridge is
\begin{equation}
\mu_{\mathrm{int}}(t;f)
\;\longleftrightarrow\;
\text{pace-dependent entropy production, damage, and repair demand},
\end{equation}
and
\begin{equation}
\mu_{\mathrm{ext}}
\;\longleftrightarrow\;
\text{ecological exposure per unit calendar time}.
\end{equation}
PBTE couples these because a change in \(f\) changes both internal pace and
external exposure duration through \(L=\Nstar/f\). This is the mathematical reason
why ecological danger and physiological pace cannot be separated in life-history
evolution.

Real species, however, do not optimize in isolation. A prey species evolves its
tempo in relation to predators; a predator evolves its tempo in relation to prey;
a host evolves its immune and metabolic schedules in relation to pathogens; a
parasite evolves replication timing in relation to host cycles. For \(n\)
interacting species, let the performance of species \(i\) be
\begin{equation}
P_i\!\left(
f_i,L_i;\mathbf{f}_{-i},\mathbf{L}_{-i}
\right),
\qquad
f_iL_i=N_{\star,i}.
\label{eq:multispecies}
\end{equation}
Here \(\mathbf{f}_{-i}\) and \(\mathbf{L}_{-i}\) denote the paces and lifespans of
all species other than \(i\). A natural decomposition is
\begin{equation}
P_i
=
\Xi_i(f_i,L_i;E)
\,
U_i(\rho_{i1},\ldots,\rho_{in}),
\qquad
\rho_{ij}=\frac{f_i}{f_j},
\label{eq:decomposition}
\end{equation}
where \(\Xi_i\) is the intrinsic demographic or thermodynamic performance of
species \(i\), and \(U_i\) describes ecological interactions through relative
tempo. The ratio \(\rho_{ij}\) measures how quickly species \(i\)'s internal time
runs relative to species \(j\)'s. This ratio is central for predator--prey timing,
host--pathogen persistence, pollination windows, developmental synchrony, and
seasonal interaction.

A PBTE--Nash equilibrium is a profile
\[
\{f_i^\ast,L_i^\ast\}_{i=1}^n
\]
such that no species can improve its own performance by unilaterally moving along
its own PBTE manifold while all other species remain fixed:
\begin{equation}
P_i\!\left(
f_i^\ast,L_i^\ast;
\mathbf{f}_{-i}^\ast,\mathbf{L}_{-i}^\ast
\right)
\ge
P_i\!\left(
f_i,L_i;
\mathbf{f}_{-i}^\ast,\mathbf{L}_{-i}^\ast
\right),
\qquad
\forall\, f_iL_i=N_{\star,i}.
\label{eq:nash}
\end{equation}
The first-order condition can be written in logarithmic form. Let
\(u_i=\log f_i\), so that \(\log L_i=\log N_{\star,i}-u_i\). Then
\begin{equation}
\frac{\dd}{\dd u_i}\log P_i
=
\left(E_{f_i}^{\Xi}-E_{L_i}^{\Xi}\right)
+
\sum_{j\ne i}
E_{\rho_{ij}}^{U},
\label{eq:multispecies_derivative}
\end{equation}
where
\begin{equation}
E_{\rho_{ij}}^{U}
=
\frac{\partial\log U_i}{\partial\log\rho_{ij}} .
\label{eq:interaction_elasticity}
\end{equation}
At equilibrium,
\begin{equation}
E_{f_i}^{\Xi}
+
\sum_{j\ne i}E_{\rho_{ij}}^{U}
=
E_{L_i}^{\Xi}.
\label{eq:pbte_nash_elasticity}
\end{equation}
This is the multi-species elasticity-balance law. In isolation, the intrinsic
elasticity of pace must equal the elasticity of lifespan. In an ecological
community, the value of pace is modified by interaction elasticities. If predator
avoidance, prey capture, immune escape, infection timing, or mutualistic
synchrony rewards a change in relative tempo, the equilibrium pace shifts even if
the intrinsic life-history surface is unchanged.

The implication is important for ecology. The diversity of tempos in an ecosystem
is not merely residual scatter around a single optimal pace. It can be a coupled
equilibrium in which each species' best position on its own PBTE manifold depends
on the temporal positions of the others. Predator--prey cycles, host--pathogen
resonance, pollinator--plant timing, microbial turnover, and successional dynamics
can therefore be interpreted as interactions among biological clocks. The
fitness-surface formalism developed here provides the bridge from individual PBTE
optimization to the ecological synchronization theory developed in the next
section.

Several empirical applications follow directly. First, for a single clade, one can
estimate demographic quantities \(b(t)\), \(S(t)\), and \(\mu(t)\), reconstruct the
reduced surface \(\widetilde P(f)\), and test whether observed species occupy
points near the predicted elasticity balance. Second, across environments, the
scaling \(f^\ast\propto\mu_{\mathrm{ext}}^{\zeta/\nu}\) predicts that lineages in
high-risk habitats should operate at higher pace and shorter lifespan after
controlling for body mass and phylogeny. Third, conservation loss of slow-lived
species should be understood not only as loss of biomass or taxonomic diversity,
but as loss of low-frequency temporal structure from the community. Fourth,
host--pathogen systems should be analyzed by tempo ratios \(f_V/f_H\), because
persistence and latency depend on whether pathogen replication lies inside or
outside the host's temporal response window. Finally, in aging and medicine, the
same operational surfaces imply that interventions cannot be evaluated solely by
whether they extend chronological lifespan; they must be classified by whether
they alter intrinsic pace, entropy cost per tick, extrinsic hazard exposure,
repair curvature, or the effective PBTE budget.

%====================================================================
\section{Ecological Synchronization}
\label{sec:eco}
%====================================================================

If each organism carries an intrinsic PBTE clock, then an ecosystem is not merely
a network of biomass, trophic exchange, population abundance, and energetic
flux. It is also a spectrum of interacting biological times. Predator--prey
cycles, pollination windows, host--pathogen persistence, circadian feeding,
seasonal reproduction, microbial turnover, and successional recovery all depend
on whether the relevant biological clocks overlap, drift apart, or become
entrained. The purpose of this section is to formulate this idea mathematically.
The central claim is not that PBTE replaces classical population ecology, but
that it supplies an additional temporal coordinate on which ecological coupling
takes place.

For species \(i\), let \(f_i(t)\) denote an effective intrinsic biological
frequency measured in cycles per unit chronological time. This frequency may
represent cardiac cycling, respiratory cycling, reproductive turnover, cellular
division, metabolic cycling, circadian activity, or another coarse-grained
biological rhythm appropriate to the ecological interaction under study. The
PBTE cycle count is
\begin{equation}
\vartheta_i(t)
=
\int_0^t f_i(s)\,\mathrm{d}s .
\label{eq:eco_cycle_count}
\end{equation}
Here \(\vartheta_i(t)\) counts how many internal biological cycles species \(i\)
has accumulated by chronological time \(t\). It is dimensionless if \(f_i\) is
measured in cycles per unit time. For synchronization theory it is often more
convenient to work with an angular phase,
\begin{equation}
\theta_i(t)
=
2\pi \vartheta_i(t)
=
2\pi\int_0^t f_i(s)\,\mathrm{d}s ,
\label{eq:eco_phase}
\end{equation}
so that one completed biological cycle corresponds to an increase of \(2\pi\) in
phase. The associated angular frequency is therefore
\begin{equation}
\omega_i(t)
=
\dot{\theta}_i(t)
=
2\pi f_i(t).
\label{eq:omega_f_relation}
\end{equation}
Equations~\eqref{eq:eco_cycle_count}--\eqref{eq:omega_f_relation} separate two
closely related but distinct quantities: \(\vartheta_i\) counts biological
cycles, while \(\theta_i\) is the angular phase used to describe synchrony,
entrainment, and phase locking.

When two biological clocks are approximately stationary over the ecological
interval of interest, their phases are
\begin{equation}
\theta_i(t)
\simeq
\theta_i(0)+\omega_i t,
\qquad
\theta_j(t)
\simeq
\theta_j(0)+\omega_j t.
\label{eq:stationary_phases}
\end{equation}
The relative phase is
\begin{equation}
\phi_{ij}(t)=\theta_i(t)-\theta_j(t),
\label{eq:relative_phase_def}
\end{equation}
and differentiating gives
\begin{equation}
\dot{\phi}_{ij}(t)
=
\dot{\theta}_i(t)-\dot{\theta}_j(t)
=
\omega_i-\omega_j .
\label{eq:relative_phase_drift}
\end{equation}
Thus, in the absence of coupling, two biological clocks drift apart at a rate
equal to their frequency mismatch. If \(|\omega_i-\omega_j|\) is large, the
relative phase rapidly sweeps through all values and persistent temporal
alignment is difficult. If \(|\omega_i-\omega_j|\) is small, even weak ecological
coupling may be sufficient to maintain a stable phase relationship.

The same idea extends to rational synchronization. Many ecological interactions
do not require one-to-one matching of cycles. A predator may attack once every
several prey reproductive cycles; a parasite may replicate many times within one
host immune cycle; a pollinator may visit during a narrow phase of plant
flowering. Thus, interaction may occur when the clocks satisfy a \(p:q\)
commensurability relation:
\begin{equation}
q\omega_i \approx p\omega_j,
\qquad p,q\in\mathbb{N}.
\label{eq:pq_resonance_angular}
\end{equation}
Using \(\omega=2\pi f\), this becomes
\begin{equation}
\frac{f_i}{f_j}
\approx
\frac{p}{q}.
\label{eq:pq_resonance_frequency}
\end{equation}
Equation~\eqref{eq:pq_resonance_frequency} is the ecological resonance condition.
It states that interaction strength is governed not only by absolute speed, but
by relative biological pace. A host and pathogen, predator and prey, or plant and
pollinator interact most effectively when their relevant internal cycles overlap
within a finite temporal window.

A natural way to quantify temporal mismatch is the logarithmic distance
\begin{equation}
d_{ij}
=
\left|
\log f_i-\log f_j
\right|
=
\left|
\log\left(\frac{f_i}{f_j}\right)
\right|.
\label{eq:log_temporal_distance}
\end{equation}
The logarithmic form is appropriate because biological frequencies often span
orders of magnitude. It also makes the PBTE relation especially transparent:
if \(f_iL_i=N_\star\), then
\begin{equation}
\log f_i+\log L_i=\log N_\star ,
\label{eq:pbte_log_line}
\end{equation}
so PBTE appears as an approximately linear constraint in
\((\log f,\log L)\) space. For \(p:q\) resonance, the corresponding mismatch is
\begin{equation}
d_{ij}^{(p:q)}
=
\left|
\log\left(\frac{qf_i}{pf_j}\right)
\right|.
\label{eq:pq_log_distance}
\end{equation}
Strong direct ecological coupling is expected when the mismatch lies inside an
interaction corridor,
\begin{equation}
d_{ij}^{(p:q)}\le d_c ,
\label{eq:interaction_corridor_general}
\end{equation}
where \(d_c\) is the maximum tolerated mismatch for effective entrainment,
exploitation, mutualism, infection, or regulation. Equivalently,
\begin{equation}
e^{-d_c}
\le
\frac{qf_i}{pf_j}
\le
e^{d_c}.
\label{eq:corridor_ratio}
\end{equation}
Equation~\eqref{eq:corridor_ratio} shows that the corridor is multiplicative,
not additive. This is important because a fixed additive difference in frequency
has very different biological meaning for slow organisms and fast organisms,
whereas a fixed ratio has comparable meaning across scales.

%--------------------------------------------------------------------
\subsection{Phase Reduction and Ecological Entrainment}
\label{subsec:phase_reduction}
%--------------------------------------------------------------------

The mathematical structure of ecological synchronization can be made explicit by
using phase-reduction theory. Suppose that each species, population, or
physiological subsystem possesses a stable biological rhythm. The full state of
subsystem \(i\) may be high-dimensional, but if the rhythm is stable, its long-time
dynamics can be described by a phase variable \(\theta_i\). Weak interactions
between such rhythms generically lead to phase equations of the form
\begin{equation}
\dot{\theta}_i
=
\omega_i
+
\sum_{j=1}^{n}
K_{ij}H_{ij}(\theta_j-\theta_i),
\label{eq:general_phase_model}
\end{equation}
where \(\omega_i\) is the natural angular frequency, \(K_{ij}\) is the effective
coupling strength from clock \(j\) to clock \(i\), and \(H_{ij}\) is a periodic
coupling function. When the coupling is weak and the leading harmonic dominates,
one obtains the Kuramoto form
\begin{equation}
\dot{\theta}_i
=
\omega_i
+
\sum_{j=1}^{n}
K_{ij}\sin(\theta_j-\theta_i).
\label{eq:eco_kuramoto}
\end{equation}
This equation has a direct ecological interpretation. The first term,
\(\omega_i\), is the intrinsic PBTE pace of species \(i\). The second term
represents temporal adjustment due to ecological interaction. If
\(\theta_j-\theta_i>0\), then clock \(j\) is ahead of clock \(i\), and the sign of
the sine term determines whether the interaction accelerates or delays clock
\(i\). Thus coupling does not merely change population abundance; it changes
phase.

The collective degree of synchronization is measured by the complex order
parameter
\begin{equation}
re^{i\psi}
=
\frac{1}{n}
\sum_{j=1}^{n}e^{i\theta_j}.
\label{eq:order_parameter}
\end{equation}
The modulus \(r\) satisfies \(0\le r\le 1\). If phases are uniformly scattered
around the circle, the complex exponentials cancel and \(r\simeq0\). If the phases
are aligned, the exponentials add coherently and \(r\simeq1\). The angle \(\psi\)
is the mean ecological phase. Thus \(r\) measures the coherence of the temporal
assemblage, while \(\psi\) identifies its collective phase.

For all-to-all coupling, \(K_{ij}=K/n\), Eq.~\eqref{eq:eco_kuramoto} becomes
\begin{equation}
\dot{\theta}_i
=
\omega_i
+
\frac{K}{n}
\sum_{j=1}^{n}\sin(\theta_j-\theta_i).
\label{eq:all_to_all_kuramoto}
\end{equation}
Using the identity
\begin{equation}
r\sin(\psi-\theta_i)
=
\frac{1}{n}
\sum_{j=1}^{n}\sin(\theta_j-\theta_i),
\label{eq:order_identity}
\end{equation}
we obtain the mean-field equation
\begin{equation}
\dot{\theta}_i
=
\omega_i
+
Kr\sin(\psi-\theta_i).
\label{eq:mean_field_kuramoto}
\end{equation}
The term \(Kr\sin(\psi-\theta_i)\) shows that each clock is pulled toward the
collective phase \(\psi\), with an effective strength proportional to both the
coupling \(K\) and the existing coherence \(r\). Synchronization is therefore
self-reinforcing: once partial coherence appears, the mean field becomes stronger.

If the natural frequencies are sampled from a symmetric unimodal distribution
\(g(\omega)\), the incoherent state loses stability at the classical threshold
\begin{equation}
K_c
=
\frac{2}{\pi g(0)},
\label{eq:kuramoto_threshold}
\end{equation}
after transforming to a frame rotating with the central frequency. This expression
has a clear PBTE interpretation. The value \(g(0)\) is large when many species
have intrinsic biological frequencies near the community mean. In that case,
only weak coupling is required to synchronize them. If the temporal spectrum is
broad, \(g(0)\) is small, and stronger coupling is required. Therefore PBTE
predicts that communities with compact biological-frequency spectra should
entrain more easily than communities whose biological clocks are widely dispersed.

%--------------------------------------------------------------------
\subsection{Two-Clock Locking and the Entrainment Corridor}
\label{subsec:two_clock_locking}
%--------------------------------------------------------------------

The essential locking mechanism is already visible in the two-clock case. Consider
two interacting biological clocks:
\begin{equation}
\dot{\theta}_1
=
\omega_1
+
K\sin(\theta_2-\theta_1),
\label{eq:theta1_two_clock}
\end{equation}
\begin{equation}
\dot{\theta}_2
=
\omega_2
+
K\sin(\theta_1-\theta_2).
\label{eq:theta2_two_clock}
\end{equation}
Define the relative phase
\begin{equation}
\phi
=
\theta_1-\theta_2 .
\label{eq:relative_phase_two_clock}
\end{equation}
Subtracting Eq.~\eqref{eq:theta2_two_clock} from Eq.~\eqref{eq:theta1_two_clock}
gives
\begin{align}
\dot{\phi}
&=
\dot{\theta}_1-\dot{\theta}_2
\nonumber\\
&=
\omega_1-\omega_2
+
K\sin(\theta_2-\theta_1)
-
K\sin(\theta_1-\theta_2).
\label{eq:phi_derivation_step1}
\end{align}
Since \(\theta_2-\theta_1=-\phi\), and since \(\sin(-\phi)=-\sin\phi\), this becomes
\begin{align}
\dot{\phi}
&=
\Delta\omega
+
K\sin(-\phi)
-
K\sin\phi
\nonumber\\
&=
\Delta\omega
-
2K\sin\phi,
\label{eq:adler_equation}
\end{align}
where
\begin{equation}
\Delta\omega=\omega_1-\omega_2 .
\label{eq:delta_omega_def}
\end{equation}
Equation~\eqref{eq:adler_equation} is the Adler equation. It states that relative
phase is driven by two competing effects: intrinsic frequency mismatch
\(\Delta\omega\), which causes drift, and coupling \(2K\sin\phi\), which resists
drift.

Phase locking occurs when \(\dot{\phi}=0\). Therefore,
\begin{equation}
\Delta\omega-2K\sin\phi^\ast=0,
\label{eq:locking_condition_step}
\end{equation}
or
\begin{equation}
\sin\phi^\ast
=
\frac{\Delta\omega}{2K}.
\label{eq:locked_sine}
\end{equation}
A real solution exists only when the right-hand side lies between \(-1\) and
\(1\). Hence the necessary and sufficient condition for phase locking is
\begin{equation}
|\Delta\omega|\le 2K.
\label{eq:locking_inequality}
\end{equation}
This is the mathematical expression of an ecological entrainment corridor.
Coupling can compensate for biological-frequency mismatch only up to a finite
threshold. If the clocks are too far apart, they cannot remain phase locked.

The locked phases are
\begin{equation}
\phi^\ast_s
=
\sin^{-1}\left(\frac{\Delta\omega}{2K}\right),
\label{eq:stable_locked_phase}
\end{equation}
and
\begin{equation}
\phi^\ast_u
=
\pi-
\sin^{-1}\left(\frac{\Delta\omega}{2K}\right).
\label{eq:unstable_locked_phase}
\end{equation}
To determine stability, define
\begin{equation}
F(\phi)
=
\Delta\omega-2K\sin\phi .
\label{eq:F_phi}
\end{equation}
Then
\begin{equation}
F'(\phi)
=
-2K\cos\phi .
\label{eq:F_prime}
\end{equation}
A locked phase is stable when \(F'(\phi^\ast)<0\), which requires
\begin{equation}
\cos\phi^\ast>0.
\label{eq:stability_condition}
\end{equation}
Therefore the stable branch is the one with \(-\pi/2<\phi^\ast<\pi/2\). At this
stable locked phase, both clocks rotate with the same common angular frequency:
\begin{equation}
\Omega
=
\dot{\theta}_1
=
\dot{\theta}_2.
\label{eq:common_frequency_def}
\end{equation}
Using Eq.~\eqref{eq:locked_sine} in Eq.~\eqref{eq:theta1_two_clock},
\begin{align}
\Omega
&=
\omega_1
+
K\sin(\theta_2-\theta_1)
\nonumber\\
&=
\omega_1
-
K\sin\phi^\ast
\nonumber\\
&=
\omega_1
-
\frac{\Delta\omega}{2}
\nonumber\\
&=
\frac{\omega_1+\omega_2}{2}.
\label{eq:common_frequency}
\end{align}
Thus, under symmetric coupling, the locked pair evolves at the arithmetic mean of
the two natural frequencies. Biologically, the faster clock is slowed, the slower
clock is accelerated, and the coupled pair settles into a shared temporal rhythm.

%--------------------------------------------------------------------
\subsection{Predator--Prey Cycles as PBTE-Coupled Clocks}
\label{subsec:predator_prey_pbte}
%--------------------------------------------------------------------

Classical population cycles provide a direct bridge between ecological dynamics
and PBTE clocks. Consider the Lotka--Volterra system
\begin{equation}
\dot{x}
=
ax-bxy,
\label{eq:lv_prey}
\end{equation}
\begin{equation}
\dot{y}
=
-dy+cxy,
\label{eq:lv_predator}
\end{equation}
where \(x(t)\) is prey density, \(y(t)\) is predator density, \(a\) is the prey
growth rate, \(d\) is the predator mortality rate, and \(b,c\) measure the
interaction strengths. The interior fixed point is obtained by setting
\(\dot{x}=0\) and \(\dot{y}=0\). From Eq.~\eqref{eq:lv_prey},
\begin{equation}
x(a-by)=0,
\label{eq:lv_fixed_step1}
\end{equation}
so for \(x>0\),
\begin{equation}
y^\ast=\frac{a}{b}.
\label{eq:lv_ystar}
\end{equation}
From Eq.~\eqref{eq:lv_predator},
\begin{equation}
y(-d+cx)=0,
\label{eq:lv_fixed_step2}
\end{equation}
so for \(y>0\),
\begin{equation}
x^\ast=\frac{d}{c}.
\label{eq:lv_xstar}
\end{equation}

To obtain the local oscillation frequency, write small deviations from the fixed
point as
\begin{equation}
x=x^\ast+\xi,
\qquad
y=y^\ast+\eta .
\label{eq:lv_deviations}
\end{equation}
The Jacobian matrix is
\begin{equation}
J
=
\begin{pmatrix}
\frac{\partial}{\partial x}(ax-bxy) &
\frac{\partial}{\partial y}(ax-bxy)
\\[4pt]
\frac{\partial}{\partial x}(-dy+cxy) &
\frac{\partial}{\partial y}(-dy+cxy)
\end{pmatrix}.
\label{eq:lv_jacobian_general}
\end{equation}
Evaluating the partial derivatives gives
\begin{equation}
J
=
\begin{pmatrix}
a-by & -bx
\\[4pt]
cy & -d+cx
\end{pmatrix}.
\label{eq:lv_jacobian_before_fixed}
\end{equation}
At the fixed point,
\begin{equation}
a-by^\ast=0,
\qquad
-d+cx^\ast=0,
\label{eq:lv_fixed_cancellations}
\end{equation}
so
\begin{equation}
J^\ast
=
\begin{pmatrix}
0 & -bx^\ast
\\[4pt]
cy^\ast & 0
\end{pmatrix}
=
\begin{pmatrix}
0 & -bd/c
\\[4pt]
ca/b & 0
\end{pmatrix}.
\label{eq:lv_jacobian_fixed}
\end{equation}
The eigenvalues satisfy
\begin{equation}
\det(J^\ast-\lambda I)=0.
\label{eq:lv_eigencondition}
\end{equation}
Explicitly,
\begin{equation}
\det
\begin{pmatrix}
-\lambda & -bd/c
\\[4pt]
ca/b & -\lambda
\end{pmatrix}
=
\lambda^2+ad
=
0.
\label{eq:lv_eigen_derivation}
\end{equation}
Therefore
\begin{equation}
\lambda_\pm
=
\pm i\sqrt{ad}.
\label{eq:lv_eigenvalues}
\end{equation}
The small-oscillation angular frequency is
\begin{equation}
\omega_0
=
\sqrt{ad},
\label{eq:lv_frequency}
\end{equation}
and the corresponding ecological period is
\begin{equation}
T_0
=
\frac{2\pi}{\sqrt{ad}}.
\label{eq:lv_period}
\end{equation}

PBTE enters when the demographic rates \(a\) and \(d\) are related to intrinsic
biological frequencies. Suppose
\begin{equation}
a=\alpha f_{\mathrm{prey}},
\qquad
d=\delta f_{\mathrm{pred}},
\label{eq:lv_pbte_rate_scaling}
\end{equation}
where \(\alpha\) and \(\delta\) are dimensionless proportionality factors encoding
ecological and physiological conversion efficiencies. Then
\begin{equation}
\omega_0
=
\sqrt{\alpha\delta f_{\mathrm{prey}}f_{\mathrm{pred}}},
\label{eq:lv_pbte_frequency}
\end{equation}
and
\begin{equation}
T_0
=
\frac{2\pi}
{\sqrt{\alpha\delta f_{\mathrm{prey}}f_{\mathrm{pred}}}}.
\label{eq:lv_pbte_period}
\end{equation}
Thus, the ecological oscillation frequency is proportional to the geometric mean
of the prey and predator biological clocks. The population cycle is not governed
by either species alone. It is a coupled temporal mode produced by both internal
rates.

This result also reveals what each species experiences internally during one
ecological cycle. The prey accumulates
\begin{align}
N_{\mathrm{prey}}^{\mathrm{cycle}}
&=
f_{\mathrm{prey}}T_0
\nonumber\\
&=
\frac{2\pi}{\sqrt{\alpha\delta}}
\sqrt{\frac{f_{\mathrm{prey}}}{f_{\mathrm{pred}}}},
\label{eq:prey_cycles_per_ecological_cycle}
\end{align}
while the predator accumulates
\begin{align}
N_{\mathrm{pred}}^{\mathrm{cycle}}
&=
f_{\mathrm{pred}}T_0
\nonumber\\
&=
\frac{2\pi}{\sqrt{\alpha\delta}}
\sqrt{\frac{f_{\mathrm{pred}}}{f_{\mathrm{prey}}}}.
\label{eq:pred_cycles_per_ecological_cycle}
\end{align}
Multiplying these two expressions gives
\begin{equation}
N_{\mathrm{prey}}^{\mathrm{cycle}}
N_{\mathrm{pred}}^{\mathrm{cycle}}
=
\frac{4\pi^2}{\alpha\delta}.
\label{eq:cycle_product}
\end{equation}
The product is independent of the frequency ratio. This provides a PBTE
interpretation of predator--prey oscillations: frequency mismatch redistributes
internal biological cycles between predator and prey, but the coupled ecological
mode preserves a constrained temporal product determined by interaction
efficiencies.

%--------------------------------------------------------------------
\subsection{Host--Pathogen Pacing}
\label{subsec:host_pathogen_pacing}
%--------------------------------------------------------------------

Host--pathogen systems provide an especially important example because the
pathogen and host can occupy dramatically different biological timescales. Let
\(f_V\) denote the effective pathogen replication frequency and let \(f_H\)
denote a host immune, physiological, or cellular-response frequency. Define the
dimensionless pacing ratio
\begin{equation}
\Gamma
=
\frac{f_V}{f_H}.
\label{eq:pacing_ratio}
\end{equation}
The ratio \(\Gamma\) measures how rapidly the pathogen advances through its
replicative internal time relative to the host's regulatory time. If
\(\Gamma\gg1\), pathogen replication is fast relative to host response. If
\(\Gamma\ll1\), pathogen replication is slow relative to host regulation.

A minimal hypothesis is that persistence is maximal not at arbitrarily large or
small \(\Gamma\), but inside a finite resonance corridor. This can be modeled by
a log-Gaussian persistence functional,
\begin{equation}
\Pi(\Gamma)
=
\Pi_0
\exp\left[
-
\frac{
\left(\log\Gamma-\log\Gamma^\ast\right)^2
}
{2\sigma_\Gamma^2}
\right],
\label{eq:persistence_functional}
\end{equation}
where \(\Pi_0\) is the maximum persistence, \(\Gamma^\ast\) is the optimal pacing
ratio, and \(\sigma_\Gamma\) is the width of the persistence corridor. The use of
\(\log\Gamma\) reflects the fact that replication and immune-response frequencies
can differ by orders of magnitude. Equation~\eqref{eq:persistence_functional}
has a maximum when
\begin{equation}
\frac{\mathrm{d}\Pi}{\mathrm{d}\log\Gamma}=0.
\label{eq:persistence_derivative_condition}
\end{equation}
Since
\begin{equation}
\frac{\mathrm{d}\log\Pi}{\mathrm{d}\log\Gamma}
=
-
\frac{\log\Gamma-\log\Gamma^\ast}{\sigma_\Gamma^2},
\label{eq:log_persistence_derivative}
\end{equation}
the maximum occurs at
\begin{equation}
\Gamma=\Gamma^\ast.
\label{eq:optimal_gamma}
\end{equation}
Thus, persistence is highest when pathogen and host clocks are neither too
separated nor too similar, but optimally matched for sustained interaction.

A therapeutic or environmental control \(u(t)\) can deform the pacing ratio:
\begin{equation}
\Gamma(t;u)
=
\frac{f_V(t;u)}{f_H(t;u)}.
\label{eq:controlled_gamma}
\end{equation}
Differentiating the logarithm gives
\begin{equation}
\frac{\mathrm{d}}{\mathrm{d}t}\log\Gamma(t;u)
=
\frac{\dot{f}_V(t;u)}{f_V(t;u)}
-
\frac{\dot{f}_H(t;u)}{f_H(t;u)}.
\label{eq:gamma_log_derivative}
\end{equation}
Therefore interventions can act by slowing pathogen internal time, accelerating
or reorganizing host response time, or shifting their relative phase. Antivirals
primarily reduce \(f_V\). Fever and metabolic stress may alter \(f_H\). Circadian
misalignment may change the phase of host defense. Block-and-lock strategies may
push the pathogen toward an effectively low-\(\Gamma\) state. In PBTE language,
therapy can be interpreted not only as reducing pathogen load, but as controlling
relative biological time.

%====================================================================
\section{Ecosystem Entropy Throughput and Temporal Resilience}
\label{sec:ecosystem}
%====================================================================

The synchronization picture describes how biological clocks align or drift. The
thermodynamic picture asks how their irreversible activity aggregates. For
species \(i\), let \(n_i(t)\) denote abundance, \(\sigma_{0,i}(t)\) the entropy
cost per biological cycle, and \(f_i(t)\) the intrinsic biological frequency.
The entropy-production rate of that population is
\begin{equation}
\dot{\Sigma}_i(t)
=
n_i(t)\sigma_{0,i}(t)f_i(t).
\label{eq:species_entropy_rate}
\end{equation}
The dimensions are transparent:
\begin{equation}
[n_i]\times
\left[\frac{\mathrm{entropy}}{\mathrm{cycle}}\right]
\times
\left[\frac{\mathrm{cycle}}{\mathrm{time}}\right]
=
\left[\frac{\mathrm{entropy}}{\mathrm{time}}\right].
\label{eq:entropy_units}
\end{equation}
Thus, abundance, entropy cost per tick, and biological pace together determine
the irreversible throughput of the population.

Summing over all species gives the ecosystem entropy-throughput rate
\begin{equation}
\dot{\Sigma}_{\mathrm{eco}}(t)
=
\sum_i n_i(t)\sigma_{0,i}(t)f_i(t).
\label{eq:ecosystem_entropy_rate}
\end{equation}
Integrating over an ecological observation interval \(T\) gives
\begin{equation}
\Sigma_{\mathrm{eco}}(T)
=
\int_0^T
\sum_i n_i(t)\sigma_{0,i}(t)f_i(t)
\,\mathrm{d}t.
\label{eq:ecosystem_entropy_total}
\end{equation}
Equation~\eqref{eq:ecosystem_entropy_total} is an ecosystem-level PBTE accounting
law. It states that ecosystem dissipation is not determined by biomass alone. The
same biomass can have very different temporal and thermodynamic roles depending
on whether it is carried by fast-turnover microbes, annual plants, insects, small
mammals, or long-lived vertebrates.

It is useful to define the fractional entropy-throughput weight of species \(i\):
\begin{equation}
w_i(t)
=
\frac{
n_i(t)\sigma_{0,i}(t)f_i(t)
}
{
\sum_j n_j(t)\sigma_{0,j}(t)f_j(t)
}.
\label{eq:entropy_weight}
\end{equation}
By construction,
\begin{equation}
\sum_i w_i(t)=1.
\label{eq:weights_sum}
\end{equation}
The quantity \(w_i\) measures the fraction of total ecosystem entropy throughput
carried by species \(i\). A species can have a large \(w_i\) because it is
abundant, because its entropy cost per cycle is large, or because its biological
clock runs rapidly.

The sensitivity of ecosystem entropy throughput to the biological frequency of
species \(k\) is
\begin{equation}
S_k
=
\frac{\partial\log\dot{\Sigma}_{\mathrm{eco}}}
{\partial\log f_k}.
\label{eq:frequency_sensitivity_def}
\end{equation}
If \(n_i\) and \(\sigma_{0,i}\) are held fixed, then
\begin{align}
S_k
&=
\frac{f_k}{\dot{\Sigma}_{\mathrm{eco}}}
\frac{\partial}{\partial f_k}
\left(
\sum_i n_i\sigma_{0,i}f_i
\right)
\nonumber\\
&=
\frac{f_k}{\dot{\Sigma}_{\mathrm{eco}}}
n_k\sigma_{0,k}
\nonumber\\
&=
\frac{n_k\sigma_{0,k}f_k}
{\sum_i n_i\sigma_{0,i}f_i}
\nonumber\\
&=
w_k.
\label{eq:sensitivity_equals_weight}
\end{align}
Thus, under fixed abundance and fixed entropy cost per tick, the logarithmic
sensitivity of ecosystem entropy throughput to a species' biological frequency
equals that species' entropy-throughput share.

More generally, abundance and entropy cost may themselves depend on frequency.
In that case,
\begin{equation}
S_k
=
w_k
\left[
1+
\frac{\partial\log n_k}{\partial\log f_k}
+
\frac{\partial\log\sigma_{0,k}}{\partial\log f_k}
\right].
\label{eq:general_sensitivity}
\end{equation}
This expression separates the direct clock effect from ecological feedbacks. The
first term, \(1\), is the direct effect of increasing the number of biological
ticks per unit time. The second term measures how abundance changes when the
species' pace changes. The third term measures how the thermodynamic cost per
tick changes when the pace changes. A species with large \(S_k\) is a temporal
keystone: a small change in its biological pace produces a disproportionately
large change in ecosystem entropy throughput.

This definition differs from ordinary biomass dominance. A rare but rapidly
cycling microbial species may have a large throughput sensitivity, whereas a
large, slow vertebrate may have a small instantaneous throughput but a large role
in long-term temporal stability. PBTE therefore predicts a division of ecological
labor by temporal frequency: microbes and decomposers dominate rapid entropy
flux, insects and annual plants shape seasonal turnover, and long-lived organisms
preserve low-frequency structure, memory, and resilience.

%--------------------------------------------------------------------
\subsection{Life-History Turnover Contribution}
\label{subsec:turnover_entropy}
%--------------------------------------------------------------------

The entropy-throughput law can also be written in terms of completed life
histories. Let \(B_i(t)\) be the rate at which individuals of species \(i\)
complete life histories or are replaced by new individuals. Then the number of
completed life histories during an interval \(T\) is
\begin{equation}
R_i(T)
=
\int_0^T B_i(t)\,\mathrm{d}t.
\label{eq:completed_life_histories}
\end{equation}
If one completed life history dissipates approximately
\begin{equation}
\Sigma_i^{\mathrm{life}}
=
\int_0^{L_i}
\sigma_{0,i}(t)f_i(t)\,\mathrm{d}t,
\label{eq:life_entropy_exact}
\end{equation}
and if \(\sigma_{0,i}\) is approximately constant over the lifespan, then
\begin{equation}
\Sigma_i^{\mathrm{life}}
\simeq
\sigma_{0,i}
\int_0^{L_i} f_i(t)\,\mathrm{d}t.
\label{eq:life_entropy_approx_step}
\end{equation}
Using the PBTE lifetime cycle count,
\begin{equation}
N_{\star,i}
=
\int_0^{L_i} f_i(t)\,\mathrm{d}t,
\label{eq:Nstar_life_history}
\end{equation}
we obtain
\begin{equation}
\Sigma_i^{\mathrm{life}}
\simeq
\sigma_{0,i}N_{\star,i}.
\label{eq:life_entropy_pbte}
\end{equation}
Therefore the turnover contribution to ecosystem entropy is
\begin{equation}
\Sigma_{\mathrm{eco}}^{\mathrm{turnover}}(T)
\simeq
\sum_i
R_i(T)\sigma_{0,i}N_{\star,i}.
\label{eq:turnover_entropy}
\end{equation}
This expression separates fast throughput from long-term storage. Short-lived
taxa have large \(R_i(T)\) and therefore dominate rapid turnover. Long-lived taxa
have smaller turnover counts but contribute slow temporal structure, buffering,
and delayed recovery capacity. Ecosystem function therefore depends not only on
how much energy flows through the system, but on the frequency bands through
which that energy is dissipated.

%--------------------------------------------------------------------
\subsection{Temporal Diversity and Resilience}
\label{subsec:temporal_diversity_resilience}
%--------------------------------------------------------------------

To quantify the temporal structure of a community, define
\begin{equation}
x_i=\log f_i.
\label{eq:log_frequency_x}
\end{equation}
Let \(W_i\) be a nonnegative ecological weight. Depending on the application,
\(W_i\) may represent abundance, biomass, entropy throughput, interaction degree,
or functional importance. Normalize the weights by
\begin{equation}
\tilde{W}_i
=
\frac{W_i}{\sum_j W_j}.
\label{eq:normalized_weights}
\end{equation}
The weighted mean log-frequency is
\begin{equation}
\bar{x}
=
\sum_i \tilde{W}_i x_i,
\label{eq:mean_log_frequency}
\end{equation}
and the temporal diversity is
\begin{equation}
D_T
=
\sum_i \tilde{W}_i(x_i-\bar{x})^2.
\label{eq:temporal_diversity}
\end{equation}
Equation~\eqref{eq:temporal_diversity} measures how broadly the community spans
biological timescales. A small \(D_T\) means that most species operate on similar
timescales. A large \(D_T\) means that the community contains both fast and slow
biological clocks.

Temporal diversity alone, however, is not always beneficial. A community may be
temporally broad but poorly synchronized. To quantify harmful mismatch, let
\(A_{ij}\) denote the ecological interaction matrix, where \(A_{ij}>0\) indicates
that species \(i\) and \(j\) interact. Define
\begin{equation}
\Delta_{\mathrm{mismatch}}
=
\frac{
\sum_{i<j} A_{ij}\tilde{W}_i\tilde{W}_j
\left(\log f_i-\log f_j\right)^2
}
{
\sum_{i<j} A_{ij}\tilde{W}_i\tilde{W}_j
}.
\label{eq:mismatch_measure}
\end{equation}
This quantity measures the average temporal separation among interacting species.
A large value indicates that interaction partners occupy incompatible biological
frequency bands.

A minimal PBTE resilience model is therefore
\begin{equation}
R
=
R_0+\alpha D_T-\beta\Delta_{\mathrm{mismatch}},
\qquad
\alpha,\beta>0.
\label{eq:resilience_model}
\end{equation}
Here \(R\) may be interpreted as inverse recovery time, \(R_0\) is baseline
resilience, \(\alpha D_T\) represents beneficial temporal complementarity, and
\(\beta\Delta_{\mathrm{mismatch}}\) represents harmful desynchronization. The
positive term reflects the idea that fast species contribute rapid nutrient
cycling, repair, and recolonization, whereas slow species contribute memory,
buffering, habitat structure, and long-term stability. The negative term reflects
the fact that diversity without temporal compatibility can become fragmentation.

This model reframes ecological restoration. Disturbance often removes
slow-frequency components first: old trees, long-lived vertebrates, apex
predators, deep-rooted perennials, and persistent mutualists. The remaining
system may retain high short-term productivity while losing low-frequency
stability. In PBTE terms, the community mean \(\bar{x}\) shifts upward, the slow
tail of the frequency distribution is depleted, and the temporal reservoir of the
ecosystem is weakened. Restoration should therefore not be evaluated only by
biomass recovery or species richness. It should also ask whether the lost
biological-frequency spectrum has been reconstructed.

The framework yields direct empirical predictions. First, ecosystems with broad
but coherent frequency spectra should recover more rapidly from disturbance than
temporally narrow systems with the same species richness. Second, systems lacking
slow-lived components should show high short-term turnover but reduced long-term
resilience. Third, species with large \(S_k\) should produce measurable changes
in ecosystem entropy throughput when their physiological pace, phenology, or
abundance changes. Fourth, phenological mismatch should increase
\(\Delta_{\mathrm{mismatch}}\) and reduce resilience even when taxonomic diversity
remains high. Fifth, restoration programmes that recover both fast turnover
channels and slow stabilizing clocks should outperform programmes that restore
biomass alone.

In summary, ecological synchronization extends PBTE from individual organisms to
communities. Each species carries an intrinsic biological clock; interactions
depend on whether those clocks are compatible; collective dynamics emerge when
many clocks entrain; and ecosystem resilience depends on the distribution,
coupling, and entropy throughput of those biological times. Ecology is therefore
not only the study of organisms in space and energy flow. It is also the study of
biological times interacting across scales.
\section{Evolutionary Control Dynamics}
\label{sec:control}

The previous sections identify the PBTE optimum and describe its ecological
couplings. The remaining question is dynamical: how does a population move toward,
remain near, or depart from this optimum? The answer is that evolution can be
represented as motion on the PBTE manifold. Let
\[
u=\log f,
\qquad
L=\Nstar e^{-u}.
\]
The reduced log-performance is
\begin{equation}
\Phi(u)
=
\log P(e^u,\Nstar e^{-u}).
\label{eq:Phi_control}
\end{equation}
A minimal adaptive dynamics is noisy gradient ascent,
\begin{equation}
\dot u
=
\gamma\frac{\dd\Phi}{\dd u}
+
\sqrt{2D}\,\xi(t),
\label{eq:gradflow}
\end{equation}
where \(\gamma\) is adaptive mobility, \(D\) is evolutionary noise, and \(\xi(t)\)
is standard white noise. Since
\[
\frac{\dd\Phi}{\dd u}=E_f-E_L,
\]
Eq.~\eqref{eq:gradflow} states that lineages shift toward higher pace when the
elasticity of performance with respect to pace exceeds that with respect to
lifespan, and shift toward lower pace when the reverse is true. Equilibrium occurs
at
\begin{equation}
E_f=E_L,
\label{eq:control_balance}
\end{equation}
which is exactly the elasticity-balance condition derived from constrained
optimization.

The stochastic dynamics \eqref{eq:gradflow} have stationary density
\begin{equation}
\rho_{\mathrm{st}}(u)
\propto
\exp\!\left[
\frac{\gamma}{D}\Phi(u)
\right],
\label{eq:stationary}
\end{equation}
up to normalization, when detailed-balance approximations are valid. Near a stable
optimum \(u^\ast\),
\[
\Phi(u)
\simeq
\Phi(u^\ast)
+
\frac{1}{2}\Phi''(u^\ast)(u-u^\ast)^2,
\qquad
\Phi''(u^\ast)<0.
\]
Thus the stationary distribution is approximately Gaussian with variance
\begin{equation}
\mathrm{Var}(u)
\simeq
\frac{D}{-\gamma\Phi''(u^\ast)}.
\label{eq:variance_control}
\end{equation}
The recovery rate from a perturbation is
\begin{equation}
\tau_{\mathrm{recover}}^{-1}
\sim
-\gamma\Phi''(u^\ast).
\label{eq:recovery}
\end{equation}
Equation~\eqref{eq:recovery} gives evolutionary resilience a quantitative meaning.
A sharply curved PBTE optimum resists perturbation and returns rapidly; a flat
optimum permits drift, polymorphism, and greater sensitivity to ecological change.

For discrete strategies, let \(p_k(t)\) be the fraction of the population using
strategy \(k\), with pace \(f_k\) and lifespan \(L_k=\Nstar/f_k\). The replicator
equation is
\begin{equation}
\dot p_k
=
p_k(W_k-\bar W),
\qquad
\bar W=\sum_j p_jW_j,
\label{eq:replicator}
\end{equation}
where
\begin{equation}
W_k=P(f_k,\Nstar/f_k).
\label{eq:Wk}
\end{equation}
Strategies with above-average PBTE-constrained performance increase in frequency,
whereas strategies below the mean decline. In the continuum limit, the replicator
dynamics become an adaptive flow on the same reduced surface \(\Phi(u)\). Thus the
variational optimum is not merely a static calculation; it is the attractor of
standard evolutionary dynamics once the PBTE constraint is imposed.

The most general formulation treats pace as a time-dependent control variable.
Life histories are not always executed at a constant rate. Organisms accelerate
during development, pause during diapause, suppress metabolism during torpor,
increase pace during reproduction, and alter physiological rate under stress.
Let \(f(t)\) be a controllable tempo and \(x(t)\) a physiological state satisfying
\begin{equation}
\dot x=F(x,f,t).
\label{eq:state_control}
\end{equation}
The objective is to maximize
\begin{equation}
J[f]
=
\int_0^L
R(x(t),f(t),t)\,\dd t
-
\int_0^L
C(f(t),t)\,\dd t,
\label{eq:pontryagin_obj}
\end{equation}
subject to the PBTE budget
\begin{equation}
\int_0^L f(t)\,\dd t=\Nstar.
\label{eq:budget_control}
\end{equation}
Introduce a costate \(p(t)\) and a Lagrange multiplier \(\lambda\) for the
intrinsic-time budget. The Hamiltonian is
\begin{equation}
H
=
R(x,f,t)-C(f,t)+pF(x,f,t)-\lambda f.
\label{eq:hamiltonian}
\end{equation}
Pontryagin's maximum principle gives the optimal-tempo condition
\begin{equation}
\frac{\partial H}{\partial f}
=
R_f-C_f+pF_f-\lambda
=
0.
\label{eq:pontryagin}
\end{equation}
The multiplier \(\lambda\) is again the shadow price of biological time. When the
instantaneous return \(R_f+pF_f\) exceeds the marginal cost \(C_f+\lambda\), the
organism should spend biological time rapidly. When the return falls below this
cost, the organism should conserve biological time. Thus torpor, diapause,
latency, developmental acceleration, reproductive bursts, and metabolic suppression
can all be interpreted as tempo-control solutions.

The application to medicine is immediate. A clinical intervention that lowers
physiological pace without loss of function changes the trajectory \(f(t)\) and
therefore delays traversal of the PBTE budget. An intervention that lowers entropy
cost per tick changes \(\sigma_0(t)\) and reduces entropy-weighted biological time
even if the raw cycle count is unchanged. An intervention that improves repair
changes the state dynamics \(F(x,f,t)\) and the cost term \(C\). Chronotherapy
chooses treatment phase so that pathological clocks are targeted when their
marginal vulnerability is high and host cost is low. In this way, the evolutionary
control problem and the medical control problem become the same mathematical
question: when should a living system spend, conserve, or redirect its finite
biological-time budget?

\section{Coupled Physiological Clocks and the Entropy Cost of Temporal Precision}
\label{sec:clocks}

The single-clock formulation of PBTE is a necessary first approximation, but it is
not the full biological reality. An organism does not contain one clock; it contains
a hierarchy of partially autonomous clocks. Cardiac rhythm, respiratory rhythm,
circadian phase, endocrine pulsatility, immune cycling, neural oscillation,
mitochondrial redox cycling, cellular division, and metabolic turnover each carry
their own intrinsic pace. These clocks are not merely parallel. They are coupled,
entrained, reset, and sometimes misaligned. A viable organism is therefore not a
single oscillator but a temporally coherent ensemble of oscillators. The problem of
biological time is consequently not only the problem of defining an internal clock,
but also the problem of explaining how many clocks are synchronized into one
organismal trajectory, what thermodynamic price that synchronization requires, and
what pathology means when coherence is lost.

Let the internal phases of an organism be collected into the vector
\begin{equation}
\bm{\theta}(t)=\bigl(\theta_1(t),\theta_2(t),\ldots,\theta_m(t)\bigr),
\qquad
\dot{\theta}_a(t)=f_a(t),
\label{eq:theta_vector}
\end{equation}
where \(a\) labels physiological subsystems and \(f_a(t)\) is the instantaneous
frequency of subsystem \(a\). The organismal biological time is not generally equal
to any one component \(\theta_a\). Instead, the effective internal time is a
weighted aggregate,
\begin{equation}
\Theta(t)=\sum_{a=1}^{m}w_a\theta_a(t),
\qquad
w_a\ge0,
\qquad
\sum_{a=1}^{m}w_a=1.
\label{eq:multiclock}
\end{equation}
The PBTE constraint is then imposed on the aggregate coordinate,
\begin{equation}
\Theta(L)=\Nstar.
\label{eq:aggregate_pbte}
\end{equation}
The weights \(w_a\) encode a physiological choice. In a cardiac allometric
application, the cardiac clock may dominate. In a metabolic or clinical
application, the weights may be shifted toward mitochondrial, inflammatory,
circadian, or endocrine clocks. In a neural or cognitive application, high-frequency
neural clocks may enter the aggregate. Thus Eq.~\eqref{eq:multiclock} is not a
mathematical convenience; it states that the organismal clock is an effective
coordinate obtained from a hierarchy of subsystem clocks.

A natural thermodynamic weighting is obtained by assigning each subsystem a
fraction of total entropy throughput. Let \(\sigma_a(t)\) be the entropy cost per
tick of subsystem \(a\). The entropy-production rate associated with that subsystem
is, to leading order,
\begin{equation}
\dot{\Sigma}_a^{(0)}(t)=\sigma_a(t)f_a(t).
\label{eq:uncoupled_sub_entropy}
\end{equation}
If the organism-level clock is to represent the entropy-weighted advancement of the
whole system, then a natural instantaneous choice is
\begin{equation}
w_a(t)=
\frac{\sigma_a(t)f_a(t)}
{\sum_{b=1}^{m}\sigma_b(t)f_b(t)}.
\label{eq:entropy_weights}
\end{equation}
With this definition, clocks that are rapidly ticking, highly dissipative, or
regulatorily expensive contribute more strongly to the organism's effective
biological age. This resolves an ambiguity in the scalar reduction: the aggregate
clock is not simply the arithmetic mean of physiological phases, but a
thermodynamic mean weighted by irreversible cost.

Coupling among clocks modifies this accounting. Let \(J_{ab}\) denote a flux of
entropy, information, or regulatory burden from subsystem \(a\) to subsystem \(b\),
and let \(\Xi_a\ge0\) denote genuine coupling dissipation generated by maintaining
coordination. The subsystem entropy balance may be written as
\begin{equation}
\dot{\Sigma}_a
=
\sigma_a f_a
+
\sum_b J_{ba}
-
\sum_b J_{ab}
+
\Xi_a .
\label{eq:entropy_coupling}
\end{equation}
Summing Eq.~\eqref{eq:entropy_coupling} over all subsystems gives
\begin{align}
\sum_a\dot{\Sigma}_a
&=
\sum_a\sigma_a f_a
+
\sum_{a,b}J_{ba}
-
\sum_{a,b}J_{ab}
+
\sum_a\Xi_a
\nonumber\\
&=
\sum_a\sigma_a f_a
+
\sum_a\Xi_a .
\label{eq:total_entropy_multiclock}
\end{align}
The internal exchange terms cancel exactly by index relabeling. This cancellation is
the multi-clock analogue of a conservation law: redistribution among clocks does
not by itself change the total entropy throughput of the organism. Only genuine
coupling dissipation, represented by \(\sum_a\Xi_a\), raises the total cost. The
biological implication is important. Coordinated redistribution can allow one clock
to slow while another compensates without violating the organismal budget. By
contrast, pathological misalignment requires continuous correction and therefore
increases entropy production.

The dynamical origin of clock coupling can be derived from phase reduction. Suppose
subsystem \(a\) is a weakly coupled limit-cycle oscillator with state
\(\mathbf{x}_a\),
\begin{equation}
\dot{\mathbf{x}}_a
=
\mathbf{F}_a(\mathbf{x}_a)
+
\epsilon\,\mathbf{G}_a(\mathbf{x}_a,\mathbf{x}_b),
\qquad
0<\epsilon\ll1.
\label{eq:limit_cycle}
\end{equation}
Let \(\Theta_a(\mathbf{x})\) be the asymptotic phase map of the uncoupled oscillator
and let \(\mathbf{x}_a^0(\theta)\) be its limit cycle. The infinitesimal phase
response curve is
\begin{equation}
\mathbf{Z}_a(\theta)
=
\nabla_{\mathbf{x}}\Theta_a(\mathbf{x})
\big|_{\mathbf{x}=\mathbf{x}_a^0(\theta)}.
\label{eq:prc}
\end{equation}
Taking the derivative of the phase along the perturbed trajectory gives, to first
order in \(\epsilon\),
\begin{equation}
\dot{\theta}_a
=
\omega_a
+
\epsilon\,
\mathbf{Z}_a(\theta_a)\cdot
\mathbf{G}_a(\theta_a,\theta_b).
\label{eq:phase_reduction}
\end{equation}
Averaging over the fast oscillation produces an effective coupling function,
\begin{equation}
H_a(\psi)
=
\frac{1}{2\pi}
\int_0^{2\pi}
\mathbf{Z}_a(\varphi)\cdot
\mathbf{G}_a(\varphi,\varphi+\psi)
\,\dd\varphi ,
\label{eq:H_average}
\end{equation}
so that the reduced phase equation becomes
\begin{equation}
\dot{\theta}_a
=
\omega_a+\epsilon H_a(\theta_b-\theta_a).
\label{eq:averaged_phase}
\end{equation}
For the relative phase \(\phi=\theta_b-\theta_a\), one obtains
\begin{equation}
\dot{\phi}
=
\Delta\omega+\epsilon H(\phi),
\qquad
\Delta\omega=\omega_b-\omega_a.
\label{eq:relative_phase}
\end{equation}
Keeping the leading sinusoidal harmonic \(H(\phi)=-K\sin\phi\) gives the Adler
equation,
\begin{equation}
\dot{\phi}
=
\Delta\omega-\epsilon K\sin\phi.
\label{eq:adler_clock}
\end{equation}
A locked state exists when \(\dot{\phi}=0\), hence
\begin{equation}
\sin\phi^\ast=\frac{\Delta\omega}{\epsilon K}.
\label{eq:locked_phase_condition}
\end{equation}
Therefore physiological locking is possible if and only if
\begin{equation}
|\Delta\omega|\le \epsilon K.
\label{eq:locking_condition_clock}
\end{equation}
The meaning of Eq.~\eqref{eq:locking_condition_clock} is direct: coupling must be
strong enough to overcome detuning. Atrioventricular delay, respiratory sinus
arrhythmia, cardiorespiratory coupling, SCN--peripheral entrainment, endocrine
coordination, immune--metabolic coupling, and social synchrony all reduce to this
structure at leading phase order. Failure of locking is not simply a timing error;
it is a transition from coherent internal time to drifting subsystem time.

Biological clocks are also noisy. Adding phase diffusion to the Adler equation gives
\begin{equation}
\dd\phi
=
(\Delta\omega-\epsilon K\sin\phi)\,\dd t
+
\sqrt{2D}\,\dd W_t,
\label{eq:stoch_adler}
\end{equation}
where \(D\) is phase diffusion and \(W_t\) is a Wiener process. Linearize near a
stable locked phase \(\phi^\ast\) by writing \(\phi=\phi^\ast+\delta\phi\). Since
\(\Delta\omega-\epsilon K\sin\phi^\ast=0\), the linearized dynamics are
\begin{equation}
\dd(\delta\phi)
=
-\epsilon K\cos\phi^\ast\,\delta\phi\,\dd t
+
\sqrt{2D}\,\dd W_t.
\label{eq:ou_phase}
\end{equation}
This is an Ornstein--Uhlenbeck process. Its stationary variance is
\begin{equation}
\mathrm{Var}(\phi)
=
\frac{D}{\epsilon K\cos\phi^\ast}.
\label{eq:phase_variance}
\end{equation}
Thus phase precision requires strong coupling. If coupling has an entropy cost
\begin{equation}
\dot{\Sigma}_{\mathrm{coup}}
=
\chi(\epsilon K)^2,
\qquad
\chi>0,
\label{eq:coupling_entropy_cost}
\end{equation}
then eliminating \(\epsilon K\) using Eq.~\eqref{eq:phase_variance} gives
\begin{equation}
\dot{\Sigma}_{\mathrm{coup}}
=
\chi
\left[
\frac{D}
{\mathrm{Var}(\phi)\cos\phi^\ast}
\right]^2.
\label{eq:precision_cost}
\end{equation}
Equation~\eqref{eq:precision_cost} is a central physical result: temporal precision
is thermodynamically expensive. At fixed noise strength and locked phase, reducing
phase variance by a factor of two requires a fourfold increase in coupling
dissipation. Biological timekeeping is therefore not infinitely precise because
precision is not free. The organism pays entropy to suppress phase noise.

This cost of precision is consistent with the thermodynamic uncertainty relation
for stochastic clocks. If a covariant biological current is used to estimate a time
interval \(T\), the relative timing uncertainty obeys a lower bound of the form
\begin{equation}
\frac{\mathrm{Var}[\widehat T]}
{\langle \widehat T\rangle^2}
\ge
\frac{2}{\dot{\Sigma}T},
\label{eq:tur}
\end{equation}
where \(\dot{\Sigma}\) is the entropy-production rate. Equation~\eqref{eq:tur}
states that timing accuracy is purchased with dissipation. The sinoatrial node, the
suprachiasmatic nucleus, and other high-fidelity timing subsystems are therefore
expected to be metabolically protected and strongly regulated because their
precision is expensive.

Coupling also renormalizes effective biological pace. Let the uncoupled rate of
clock \(i\) be \(f_i^{(0)}\), and suppose coupling shifts it to
\begin{equation}
\widetilde f_i
=
f_i^{(0)}+\Delta f_i(\epsilon,K,\bm{x}).
\label{eq:renormalized_rate}
\end{equation}
If the effective PBTE budget remains \(N_{\star,i}\), the corresponding
chronological lifespan is
\begin{equation}
\widetilde L_i=
\frac{N_{\star,i}}{\widetilde f_i}.
\label{eq:renormalized_lifespan}
\end{equation}
Relative to the uncoupled value
\(L_i^{(0)}=N_{\star,i}/f_i^{(0)}\),
\begin{equation}
\frac{\widetilde L_i}{L_i^{(0)}}
=
\frac{f_i^{(0)}}{\widetilde f_i}.
\label{eq:renorm}
\end{equation}
This equation is a compact bridge from physiological synchronization to healthspan.
Coordinated slowing, hibernation, metabolic coherence, symbiosis, or therapeutic
hypometabolism extend chronological duration by reducing effective pace. Chronic
inflammation, autonomic dysregulation, circadian disruption, fever, metabolic
syndrome, and proliferative disease compress duration by increasing effective pace
or entropy cost per tick. The sign of \(\Delta f_i\) distinguishes temporal
protection from temporal acceleration.

\section{Gauge Coherence of Biological Time}
\label{sec:gauge}

The gauge formulation identifies which part of a biological phase is physically
meaningful. The accumulated phase
\begin{equation}
\theta(t)=\int_0^t f(s)\,\dd s
\label{eq:theta_gauge}
\end{equation}
counts completed cycles, but its absolute origin is arbitrary. There is no
privileged zero of cardiac phase, circadian phase, respiratory phase, or endocrine
phase. This redundancy is represented by a local phase relabeling,
\begin{equation}
\theta(t)\mapsto \theta'(t)=\theta(t)+\lambda(t),
\label{eq:phase_shift}
\end{equation}
where \(\lambda(t)\) is an admissible smooth function, possibly with jumps at reset
events. The raw derivative is not invariant:
\[
\dot{\theta}'=\dot{\theta}+\dot{\lambda}.
\]
A phase reset would therefore appear as an artificial spike in physiological rate.
To eliminate this artifact, introduce a temporal connection \(A_0(t)\) transforming
as
\begin{equation}
A_0'(t)=A_0(t)+\dot{\lambda}(t).
\label{eq:connection}
\end{equation}
The covariant derivative
\begin{equation}
D_0\theta=\dot{\theta}-A_0
\label{eq:covariant_derivative}
\end{equation}
is invariant, since
\begin{equation}
D_0'\theta'
=
\dot{\theta}+\dot{\lambda}
-
(A_0+\dot{\lambda})
=
D_0\theta.
\label{eq:covariant_invariance}
\end{equation}
The physical biological rate is therefore not \(\dot{\theta}\) but
\begin{equation}
f_{\mathrm{phys}}(t)=D_0\theta(t).
\label{eq:fphys}
\end{equation}
The PBTE constraint must be written in covariant form:
\begin{equation}
\int_0^L D_0\theta\,\dd t=\Nstar.
\label{eq:pbte_covariant}
\end{equation}
This is the gauge-invariant statement of biological proper time.

The physiological meaning of \(A_0\) is concrete. In cardiac phase resetting, a
pacemaker or autonomic correction can shift the phase by \(\Delta\theta\) at time
\(t_0\). This is represented by
\[
A_0(t)=\Delta\theta\,\delta(t-t_0),
\]
so that the covariant rate \(D_0\theta\) remains finite. The reset changes the
counting convention, not the intrinsic progression. In circadian entrainment,
photic and feeding signals enter through \(A_0\); jet lag corresponds to a mismatch
among tissue-specific connections \(A_0^{(i)}\). In torpor,
\begin{equation}
D_0\theta \approx f^{\mathrm{tor}}\ll f^{\mathrm{active}},
\label{eq:torpor_rate}
\end{equation}
so biological time genuinely slows. If a hibernator spends a fraction \(q\) of its
life in torpor, the mean covariant rate is
\begin{equation}
\bar f_{\mathrm{phys}}
=
(1-q)f^{\mathrm{active}}+qf^{\mathrm{tor}},
\label{eq:mean_torpor_rate}
\end{equation}
and therefore
\begin{equation}
L\simeq \frac{\Nstar}{\bar f_{\mathrm{phys}}}.
\label{eq:torpor_lifespan}
\end{equation}
This is the covariant expression of biological time dilation. Torpor does not
violate PBTE; it reduces the rate at which the PBTE budget is spent.

A minimal gauge-invariant Lagrangian for a single biological phase is
\begin{equation}
\mathcal{L}
=
\frac{m}{2}(D_0\theta)^2
+
J_0A_0,
\label{eq:gauge_lagrangian}
\end{equation}
where \(m>0\) is an effective temporal inertia and \(J_0\) is an imposed temporal
flux. Varying with respect to \(\theta\) gives
\begin{equation}
\frac{\dd}{\dd t}\left(mD_0\theta\right)=0,
\label{eq:noether_derivation}
\end{equation}
so that
\begin{equation}
I=mD_0\theta=\mathrm{constant}.
\label{eq:noether_charge}
\end{equation}
This is the Noether charge associated with additive phase symmetry. Varying with
respect to \(A_0\) gives
\begin{equation}
-mD_0\theta+J_0=0,
\qquad
D_0\theta=\frac{J_0}{m}.
\label{eq:A0_constraint}
\end{equation}
Combining Eq.~\eqref{eq:A0_constraint} with the covariant PBTE constraint
\eqref{eq:pbte_covariant} yields
\begin{equation}
L=
\frac{\Nstar}{D_0\theta}
=
\frac{\Nstar m}{J_0}.
\label{eq:noether_lifespan}
\end{equation}
The lifespan--rate hyperbola is therefore obtained dynamically from conservation of
a temporal Noether charge under fixed covariant flux. For example, a mouse with
\(\bar f_H\approx10\) Hz and \(L\approx3\) yr and an elephant with
\(\bar f_H\approx0.5\) Hz and \(L\approx60\) yr differ by roughly a factor of
twenty in both rate and lifespan, yet accumulate the same order of biological
cycles.

For \(N\) coupled clocks sharing a common connection, the natural generalization is
\begin{equation}
\mathcal{L}
=
\sum_{i=1}^{N}
\frac{m_i}{2}(D_0\theta_i)^2
+
J_0A_0 .
\label{eq:multi_lagrangian}
\end{equation}
Variation with respect to \(A_0\) yields the global throughput constraint
\begin{equation}
\sum_{i=1}^{N}m_iD_0\theta_i=J_0.
\label{eq:global_constraint}
\end{equation}
Only relative phases
\begin{equation}
\psi_{ij}=\theta_i-\theta_j
\label{eq:relative_phase_observable}
\end{equation}
are directly observable. The weighted mean phase is gauge-dependent. This matches
physiology: we observe SCN--liver phase offset, atrioventricular delay,
respiratory--cardiac coupling, and endocrine lags, not absolute phases in any
privileged sense. When one clock slows, the global constraint determines how the
ensemble redistributes throughput. This is the gauge-theoretic version of lifespan
redistribution.

Additive phase symmetry, however, does not exhaust the geometry of biological time.
Different tissues can run at different local rates, and such differences cannot be
removed merely by shifting phase origins. This motivates a second symmetry: local
time-scale rescaling,
\begin{equation}
f(x,t)\mapsto \Lambda(x,t)f(x,t),
\qquad
\Lambda(x,t)>0.
\label{eq:weyl_transform}
\end{equation}
Introduce a Weyl connection \(\omega_\mu\) that transforms as
\begin{equation}
\omega_\mu\mapsto \omega_\mu+\partial_\mu\ln\Lambda.
\label{eq:weyl_connection}
\end{equation}
For a rate-like field \(q\) of Weyl weight \(w\), the Weyl-covariant derivative is
\begin{equation}
\nabla_\mu^{(W)}q
=
\partial_\mu q-w\omega_\mu q.
\label{eq:weyl_covariant}
\end{equation}
Unlike the purely temporal additive connection, the Weyl connection can carry
curvature,
\begin{equation}
\Omega_{\mu\nu}
=
\partial_\mu\omega_\nu-\partial_\nu\omega_\mu.
\label{eq:weyl_curvature}
\end{equation}
This curvature measures temporal incompatibility: a local mismatch of biological
time scales that cannot be eliminated by smooth relabeling. The leading
dissipation associated with such incompatibility is quadratic,
\begin{equation}
\dot{\Sigma}_{\mathrm{misalign}}
\propto
\Omega_{\mu\nu}\Omega^{\mu\nu}.
\label{eq:misalignment_cost}
\end{equation}
For small inter-tissue phase mismatch \(\Delta\phi\), this implies the empirical
form
\begin{equation}
\dot{\Sigma}
\simeq
\dot{\Sigma}_0
\left[
1+\kappa(\Delta\phi)^2
\right].
\label{eq:quadratic_mismatch}
\end{equation}
The clinical interpretation is direct: chronic circadian misalignment, shift work,
irregular feeding, sleep disruption, and tissue-specific clock desynchrony impose a
curvature cost. They raise entropy production because the organism must continually
maintain coordination among incompatible temporal gauges.

Both additive and Weyl structures can be assembled into a gauge-invariant field
action over a physiological domain:
\begin{equation}
S[\theta,A]
=
\int \dd^{d+1}x
\left[
\frac{\kappa}{2}(D_\mu\theta)(D^\mu\theta)
-
V(D_0\theta)
-
\frac{1}{4g^2}F_{\mu\nu}F^{\mu\nu}
\right],
\label{eq:field_action}
\end{equation}
with
\begin{equation}
F_{\mu\nu}=\partial_\mu A_\nu-\partial_\nu A_\mu.
\label{eq:field_strength}
\end{equation}
In this language, pathology is temporal decoherence: subsystems evolve in
incompatible gauges and produce entropy above the minimum required for coherent
function. Therapy is the restoration of a shared temporal gauge or the reduction of
curvature. Biological time is therefore a gauge-invariant quantity: arbitrary phase
labels have no physiological meaning, whereas covariant progression, relative
phase, curvature, and entropy cost do.

\section{Applications to Aging, Disease, Medicine, and Healthspan}
\label{sec:medicine}

These constructions immediately translate into medicine and healthspan. Clinical
age should be represented by the pair
\begin{equation}
\left(\Theta(t),\dot{\Theta}(t)\right),
\label{eq:clinical_pair}
\end{equation}
where \(\Theta(t)\) is accumulated internal time and \(\dot{\Theta}(t)\) is current
pace. A person may be old in accumulated biological time but currently stable; a
chronologically young patient may be undergoing rapid biological acceleration.
Diagnostics should therefore distinguish state estimators from rate estimators.
State estimators include DNA methylation age, frailty, accumulated molecular
damage, telomere attrition, proteostatic loss, and organ reserve. Rate estimators
include resting heart rate, respiratory rate, heart-rate variability, inflammatory
burden, metabolic power, mitochondrial coupling, sleep regularity, and circadian
phase coherence.

If an intervention reduces effective pace by a factor \(0<\rho<1\),
\begin{equation}
f\mapsto \rho f,
\label{eq:pace_reduction}
\end{equation}
and the available budget remains unchanged, then the PBTE prediction is
\begin{equation}
L_{\mathrm{int}}
=
\frac{\Nstar}{\rho f}
=
\frac{1}{\rho}L_{\mathrm{ctrl}}.
\label{eq:rate_reduction}
\end{equation}
This is the ideal time-dilation law. It applies only when reduced pace preserves
function. Caloric restriction, torpor, therapeutic hypometabolism, coherent
bradycardia, and temperature reduction are candidates for this class. A second
class of intervention lowers entropy cost per tick,
\begin{equation}
\sigma_0\mapsto q\sigma_0,
\qquad
0<q<1,
\label{eq:entropy_cost_reduction}
\end{equation}
so that entropy-weighted biological time accumulates more slowly even if raw cycle
count is unchanged. Improved mitochondrial coupling, reduced chronic inflammation,
enhanced proteostasis, and more efficient repair belong to this class.

Disease corresponds to the opposite deformation. Write
\begin{equation}
f(t)=f_0+\delta f(t),
\label{eq:disease_rate}
\end{equation}
so that the excess biological time consumed during a disease interval is
\begin{equation}
\Delta\theta_{\mathrm{disease}}
=
\int_{t_1}^{t_2}\delta f(t)\,\dd t.
\label{eq:excess_time}
\end{equation}
Acute infection and fever generate transient positive spikes in \(\delta f\).
Chronic inflammation creates persistent elevation. Hyperthyroidism and metabolic
syndrome accelerate biological-time velocity. Cancer can be described as local
temporal acceleration: a tumor partially decouples its internal proliferative clock
from organism-level regulation. Viral latency is the complementary state, a local
temporal arrest in which \(f_V\simeq0\) and the viral biological clock nearly stops
until reactivation. Cancer and latency are therefore opposite poles of local
time-deformation: one accelerates internal time; the other suspends it.

\subsection{Chronotherapy}
\label{sec:chronotherapy}

Chronotherapy is the controlled use of phase and rate. Let \(H(\phi,u)\) be host
damage and \(T(\phi,u)\) tumor or pathogen damage at circadian phase \(\phi\) under
intervention \(u\). The therapeutic index is
\begin{equation}
\mathcal{I}(\phi)=\frac{T(\phi,u)}{H(\phi,u)}.
\label{eq:therapeutic_index}
\end{equation}
The optimal phase is
\begin{equation}
\phi^\ast=\arg\max_\phi \mathcal{I}(\phi).
\label{eq:chrono_phase}
\end{equation}
In PBTE variables, this is phase-selective control of differential biological pace:
\begin{equation}
\phi^\ast
=
\arg\max_\phi
\left[
\frac{\Delta f_T(\phi)}{f_T}
-
\frac{\Delta f_H(\phi)}{f_H}
\right].
\label{eq:chrono_pbte}
\end{equation}
The treatment should be delivered when the pathological clock is maximally
vulnerable relative to the host clock. Chronotherapy is therefore not a timing
curiosity; it is the optimal-control problem for coupled PBTE clocks.

\subsection{Thermodynamic Habits and Preservation of Biological Time}
\label{sec:habits}

The same language gives a cautious interpretation of health habits. Sleep, morning
light, exercise, fasting rhythms, breathing practice, thermal adaptation, cognitive
engagement, and social synchrony do not stop biological time. Their PBTE
interpretation is that they reduce temporal roughness, phase variance, curvature,
or entropy cost of misalignment. Sleep repairs temporal fragmentation. Morning
light fixes a circadian boundary condition. Exercise improves metabolic
flexibility and reduces the cost of transitions among physiological regimes.
Slow breathing entrains cardiorespiratory clocks. Social synchrony acts as
distributed oscillator coupling. These habits may preserve temporal coherence, but
they should not be overclaimed as literal lifespan-maximizing mechanisms without
direct validation.

The conceptual consequence is that PBTE replaces chronological primacy with
internal-time primacy. A living system does not merely exist in time; it constructs
a thermodynamic worldline. Identity is the continuity of this worldline. Agency is
the ability of the organism to regulate its pace, phase, and coherence. Pathology
is temporal acceleration, temporal arrest, or temporal decoherence. Healthspan is
the maintenance of coherent, low-cost progression through biological time. The
invariant \(N_\star\) is not the whole theory; it is the simplest observable
projection of a deeper geometry of internally timed, entropy-producing life.
\section{Implications and Limitations}
\label{sec:implications}

The coupled-clock and gauge formulations extend PBTE from a lifetime-cycle
regularity into a theory of temporal organization. For evolution, the implication
is that life-history diversity is constrained motion on an admissible manifold:
selection may change hazard, repair, fecundity, developmental timing, and entropy
cost per tick, but it cannot remove the irreversible thermodynamic cost of
sustaining biological function. For ecology, the implication is that communities
are spectra of coupled biological clocks; resilience depends not only on species
richness or biomass but also on temporal richness, the presence of both fast repair
channels and slow stabilizing reservoirs. For medicine, the implication is that
clinical time should be two-dimensional, separating accumulated biological time
from current biological pace. Chronic inflammation, cancer, viral latency, sleep
disruption, and circadian misalignment become disorders of internal time. For
theoretical biology, the implication is that phase arbitrariness, entrainment,
resetting, and misalignment can be described geometrically: pathology is not merely
loss of function but failure to maintain a coherent temporal gauge.

The limitations are substantial and define the experimental frontier. The empirical
foundation of PBTE is strongest for endothermic vertebrates and weaker for plants,
microbes, viruses, ecosystems, and clinical intervention studies, where the
framework remains conjectural. The entropy-per-cycle closure requires direct
calorimetric validation rather than inference from proxies alone. Ecological
synchronization and gauge curvature are mathematically natural extensions, but they
must be tested independently. Biological time is multidimensional; any scalar
\(\Theta(t)\) is an approximation to a vector of coupled clocks. Many clinical
applications rely on indirect markers of entropy production, such as heart rate,
inflammatory markers, metabolic power, mitochondrial function, or circadian
misalignment. These limitations do not invalidate the framework; they identify what
must be measured for the theory to become empirical.

\section{Open Problems and Falsifiability}
\label{sec:open_problems}

The framework is deliberately formulated to be falsifiable. We list the principal
open problems whose resolution would either substantiate or refute it.
\begin{enumerate}[itemsep=2pt,topsep=2pt]
\item \emph{Choice of clock.} What is the correct physiological frequency \(f(t)\)
for each system---cardiac, respiratory, metabolic, circadian, neural, or an
entropy-weighted aggregate---and how should the aggregation weights be determined?
\item \emph{Universality of the budget.} How universal is \(N_\star\)? Within which
clades and physiological classes is it narrowly distributed, and where does it fail?
\item \emph{Measuring entropy cost.} How should the entropy cost per cycle
\(\sigma_0\) be measured experimentally, beyond the homeostatic closure
\(\sigma_0\simeq P/(Tf)\)?
\item \emph{Operational estimation.} Can entropy cost per biological tick be
estimated in practice from indirect calorimetry, body temperature, and wearable
physiology, and reconstructed as \(\sigma_0=\dot e_p/f\)?
\item \emph{Clade corrections.} How should PBTE be corrected for clade, temperature,
torpor, diving, birds, bats, primates, plants, microbes, and viruses, and are these
corrections structured multipliers \(\Phi_i\) rather than noise?
\item \emph{Predictive power.} Can biological age be predicted better by the
entropy-normalized coordinate \(A_{\mathrm{PBTE}}\) than by chronological age, out of
sample, against independent damage, frailty, and survival outcomes?
\item \emph{Cost of desynchrony.} Can desynchronization be measured as an increase
in entropy cost per useful biological tick, as the Weyl-curvature term predicts?
\item \emph{Temporal resilience.} Can temporal resilience be measured by recovery
time after a standardized perturbation, and does it scale with the curvature of the
fitness or damage landscape?
\item \emph{Gauge observables.} Can gauge curvature or inter-tissue temporal
incompatibility be operationalized in real physiological data, for example as an
SCN--peripheral phase mismatch?
\item \emph{Decisive measurement.} What single measurement would most directly test
the framework? We argue it is direct calorimetric measurement of entropy production
per physiological cycle across body mass, clade, disease state, and intervention.
\end{enumerate}

The framework would be weakened or refuted under several specific conditions. PBTE
loses support if carefully normalized lifetime internal-time budgets show no
clustering within clades after appropriate corrections; if the entropy cost per
biological tick is unconstrained across comparable physiological regimes; if
entropy-normalized internal-time accumulation fails to predict aging, frailty, or
mortality better than chronological age out of sample; if the proposed clock
variables cannot be operationally defined and reconstructed without circularity; or
if coupled physiological clocks systematically violate the predicted locking and
precision--dissipation relations. Stating these conditions explicitly is what
distinguishes a falsifiable programme from a descriptive analogy.

\section{Conclusion}
\label{sec:conclusion}

This paper has developed PBTE as a unified thermodynamic framework for biological
time. Biological proper time is defined as accumulated internal physiological
activity, \(\theta(t)=\int_0^t f(s)\,\dd s\), and its entropy-normalized extension
\(A_{\mathrm{PBTE}}\) measures the fraction of a finite reference entropy--cycle
budget already consumed. PBTE connects lifespan to this finite budget through
\(fL\approx N_\star\), and entropy production supplies the budget with
thermodynamic meaning by assigning a cost \(\sigma_0\) to each biological tick.
Evolution optimizes life history as constrained motion on the PBTE manifold, with
the scale-free condition \(E_f=E_L\) at equilibrium and a shadow price of biological
time that rises under ecological hazard. Ecological systems are spectra of
interacting biological clocks whose synchronization is governed by locking and
resonance conditions. Physiological health depends on the coherence of an ensemble
of coupled clocks, and temporal precision carries an irreducible entropy cost.
Gauge theory distinguishes the arbitrary label of biological phase from the
physically meaningful covariant rate \(D_0\theta=\dot\theta-A_0\), and recovers the
lifespan--rate relation from a conserved temporal Noether charge. Within this
framework, aging, disease, and healthspan are interpreted as changes in
internal-time rate, entropy cost per tick, or synchronization, and the construction
is testable and falsifiable rather than merely descriptive.

PBTE therefore reframes living systems as finite thermodynamic trajectories:
organisms do not merely pass through chronological time, but generate, regulate,
synchronize, and spend an internally produced biological duration.

\end{document}